\documentclass[fleqn,usenatbib]{mnras}

\usepackage{newtxtext,newtxmath, comment}

\usepackage[T1]{fontenc}

\DeclareRobustCommand{\VAN}[3]{#2}
\let\VANthebibliography\thebibliography
\def\thebibliography{\DeclareRobustCommand{\VAN}[3]{##3}\VANthebibliography}

\usepackage{graphicx}	
\usepackage{amsmath}	
\usepackage{cuted}
\usepackage{multirow} 
\usepackage{nccmath}

\title[Semi-Analytic Array Element Coupling I]{Array Element Coupling in Radio Interferometry I: A Semi-Analytic Approach}

\author[A. T. Josaitis et al.]{
Alec T. Josaitis,$^{1}$\thanks{E-mail: atj28@cam.ac.uk}
Aaron Ewall-Wice,$^{2}$
Nicolas Fagnoni,$^{1}$
Eloy de Lera Acedo$^{1,3}$
\\
$^{1}$Astrophysics Group - Cavendish Laboratory, University of Cambridge, JJ Thompson Avenue, Cambridge CB3 OHE, UK\\
$^{2}$Astronomy Department, University of California - Berkeley, 425 Campbell Hall, University Dr., Berkeley 94720, USA\\
$^{3}$Kavli Institute for Cosmology, Madingley Road, Cambridge CB3 0HA, UK
}

\date{Accepted XXX. Received YYY; in original form ZZZ}

\pubyear{2022}

\begin{document}
\label{firstpage}
\pagerange{\pageref{firstpage}--\pageref{lastpage}}
\maketitle

\begin{abstract}
We derive a general formalism for interferometric visibilities, which considers first-order antenna-antenna coupling and assumes steady-state, incident radiation. We simulate such coupling features for non-polarized skies on a compact, redundantly-spaced array and present a phenomenological analysis of the coupling features. Contrary to previous studies, we find mutual coupling features manifest themselves at nonzero fringe rates. We compare power spectrum results for both coupled and non-coupled (noiseless, simulated) data  and find coupling effects to be highly dependent on LST, baseline length, and baseline orientation. For all LSTs, lengths, and orientations, coupling features appear at delays which are outside the foreground 'wedge', which has been studied extensively and contains non-coupled astrophysical foreground features. Further, we find that first-order coupling effects threaten our ability to average data from baselines with identical length and orientation. Two filtering strategies are proposed which may mitigate such coupling systematics. The semi-analytic coupling model herein presented may be used to study mutual coupling systematics as a function of LST, baseline length, and baseline orientation. Such a model is not only helpful to the field of 21cm cosmology, but any study involving interferometric measurements, where coupling effects at the level of at least 1 part in $10^4$ could corrupt the scientific result. Our model may be used to mitigate coupling systematics in existing radio interferometers and to design future arrays where the configuration of array elements inherently mitigates coupling effects at desired LSTs and angular resolutions. 
\end{abstract}



\section{Introduction} \label{sec:INTRO}
Radio interferometers are key instruments for the fields of cosmology, astrophysics, astronomy, and ionospheric physics. The study of the 21cm emission of the hyperfine splitting of the ground state of neutral hydrogen in the intergalactic medium (IGM) is an active area of research with implications for all such disciplines. \citet{21cm_Van_de_Hulst_1945}, \citet{21cm_general_Rees_Madau}, and \citet{21cm_general_Wang} describe how a measurement of the 21cm signal probes the era of the formation of the first luminous objects (the 'Cosmic Dawn') and the subsequent phase transition of the Universe, during which photons from the first stars ionized the neutral hydrogen of the IGM at z$\approx$6 (the 'Epoch of Reionization' (EoR)). Reviews such as \citet{Furlanetto2006}, \citet{Pritchard_Loeb_Review}, \citet{LiuShaw2020}, and \citet{Mesinger_Review} carefully consider various experimental and theoretical considerations regarding a measurement of the 21cm signal, which has since been redshifted to the VHF band.

Considering that the 21cm signal is expected to be many orders of magnitude fainter than astrophysical foregrounds - not to mention the difficulties of terrestrial foregrounds, such as VHF radio frequency interference (RFI) - it is imperative that any experiment carefully mitigate instrumental systematic effects. These systematics result from a variety of sources, such as Faraday rotation due to the ionosphere, impedance mismatches throughout the analog signal chain, mischaracterization of the primary beam and/or beam sidelobes, among others. Any such systematic effect could distort the spectral structure of either the foregrounds or the EoR signal, making it more difficult to discriminate the latter from the former. \citet{Jelic2010LOFAR_foregorunds} simulate how Faraday-rotated, polarized galactic emission can affect the calibration of a radio interferometer and contaminate measurements of the 21cm signal. \citet{polarized_foregrounds} discuss how this polarized contamination of the intrinsically unpolarized 21cm power spectrum can occur at levels which are several orders of magnitude above the expected 21 cm EoR signal. \citet{Polarized_ionosphere_attenuation} describe how such leakage effects are attenuated by at least an order of magnitude due to Faraday rotation from the Earth's ionosphere, however the modelling and removal of polarized sources required to eliminate polarization leakage exceeds the reasonable capabilities of current 21cm instruments. Not only leakage from polarized foregrounds, but any form of uncalibrated structure in the instrument's frequency response may be imprinted on the foregrounds and has the potential to leak power into the  EoR  window  at  small  line  of  sight  scales,  masking the signal. \citet{Zheng2014_MITEOR} discusses how phase-switching hardware may be used to mitigate circuit-level crosstalk in the analog signal chains of a 21cm experiment. \citet{Hibbard2020} and \citet{Nithya2016_Blackman_Harris} discuss the challenges that antenna beam chromaticity poses to measurements of both the spatially-averaged ('global') 21cm signal and the angular-resolution-dependent (interferometric) 21cm power spectrum, respectively. \citet{COSMOLOGICAL_CONSTRAINTS} analyzes how a measurement of the power spectrum across a wide range of redshifts (and, analogously, a wide frequency band) offers the scientific community an understanding of how the EoR and Cosmic Dawn evolve over time, and offers independent and significantly improved constraints on many fundamental, time-independent cosmological parameters which govern the evolution of the early universe. Wideband antenna elements are likely required for such measurements. Unfortunately,  \citet{Craeye_Macro_Basis_Func_Currents} conclude that wider bandwidth antennae require larger element volumes over which to electromagnetically characterize the effective number of degrees of freedom of an antenna element; this scaling challenge makes beam chromaticity inherently more difficult to model through electromagnetic simulation. Furthermore, \citet{Craeye_Macro_Basis_Func_Currents} discuss how such a wideband system typically permits a greater number of current modes inside the antenna, which yields greater potential for chromatic variation between array elements. The wideband requirements of a 21cm measurement make the challenges of beam chromaticity inherently more difficult to electromagnetically model and constrain during the mechanical construction of a large array.

A conclusive detection of neither the global 21cm signal nor the 21cm power spectrum has yet to occur. \citet{EDGES_RESULT} describe a potential first-detection of the global signal by the EDGES collaboration, claiming to have measured an absorption trough centred at 78MHz. Subsequent examination of the data, such as that performed in \citet{Hills2018_EDGES_REBUTTAL}, indicates that the EDGES result may be corrupted by unaccounted systematic errors, which result in unphysical parameters for galactic foreground emission, and finds simple, alternative formulations for the signal which equally fit the EDGES data. \citet{SimsPober2020_EDGES_Sinusoid} show that a damped sinusoidal systematic is strongly preferred in the EDGES data. More recently, \citet{Bevins2021_MAXSMOOTH} identify a similar systematic. First-generation radio interferometers, such as the Precision Array for Probing the Epoch of Reionization (PAPER, \citet{PAPER_overview}), the Murchison Widefield Array (MWA, \citet{MWA_overview}) and the Low Frequency Array (LOFAR, \citet{LOFAR_overview}) have made noteworthy upper limits on measurements of the full 21cm power spectrum. Second-generation radio interferometers, such as the Hydrogen Epoch of Reionization Array (HERA, \citet{HERA_overview}) are currently being commissioned, and are designed to have the sensitivity to detect the 21cm power spectrum across a wide range of redshifts. 

\begin{figure}
    \centering
    \includegraphics[width=.46\textwidth]{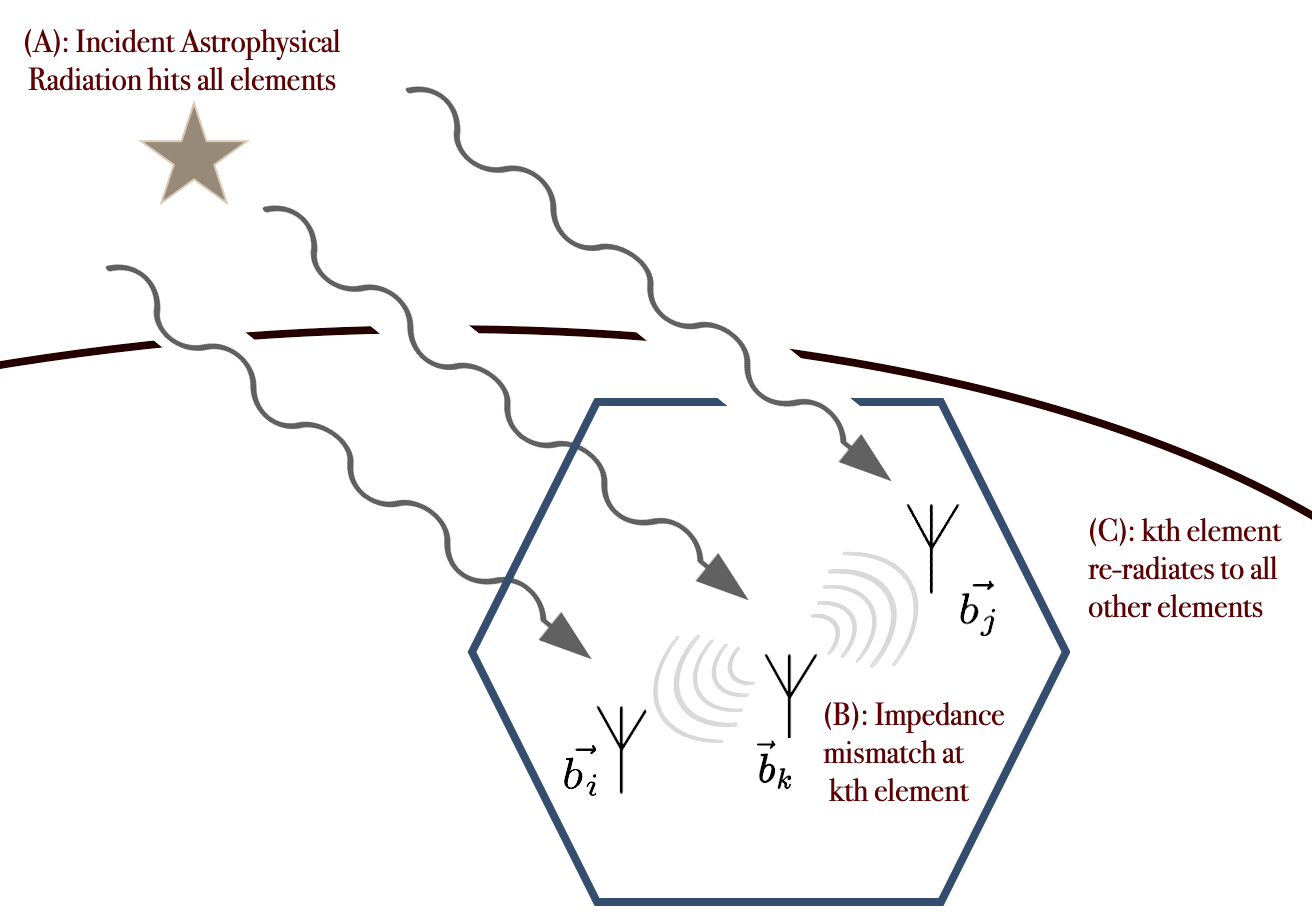}
    \caption{Conceptual outline of first-order array element coupling: Incident astrophysical radiation reaches all elements in the interferometric array. Each element in the array, being imperfectly impedance matched at its feed terminals, will reflect some amount of this incident radiation, and re-radiate power across the array, which will be absorbed by another interferometric element. Being a first-order formalism, we only study the effect of this re-radiated power being absorbed by other array elements just once (e.g. sky $\rightarrow$ antenna k $\rightarrow$ antenna i), not when that scattered power is subsequently re-radiated by all other elements (e.g. sky $\rightarrow$ antenna k  $\rightarrow$ antenna i $\rightarrow$ antenna j).  }
    \label{fig:EM_re-radiation}
\end{figure}

This work focuses on first-order antenna-antenna coupling in radio interferometers, a form of 'mutual coupling' which affects interferometric measurements of the 21cm power spectrum. Specifically, we derive a semi-analytic model of the interferometric visibility equation which considers this first-order coupling. In this model, incident astrophysical radiation enters all interferometric elements in the array. Each element absorbs most of its incident radiation, although some amount of power is then reflected due to departure from conjugate match at the terminals of the antenna feeds. Thus, each element in the array not only absorbs power from the sky but re-radiates power across the array, which is subsequently absorbed by other elements. Being a first-order formalism, we only model the effect of re-radiated power being absorbed just once more by all array elements (e.g. sky $\rightarrow$ antenna k $\rightarrow$ antenna i), not when the scattered power is subsequently absorbed and re-radiated again by all elements (e.g. sky $\rightarrow$ antenna k  $\rightarrow$ antenna i $\rightarrow$ antenna j). Figure (\ref{fig:EM_re-radiation}) provides a simplified, conceptual sketch of our first-order coupling model.

Mutual coupling systematics are a known concern for interferometric 21cm experiments (\citet{Parsons2012}, \citet{Systematics_Chaudhari}, \citet{Kern:2019}). Upper limits from all first-generation interferometers, and current limits from the the second generation HERA (which has yet to achieve its full sensitivity) report excess noise in their power spectrum measurements with a spectral correlation which is distinct from thermal noise. Analysis of all of these upper limits cite mutual coupling as one unresolved possibility for their excess power (\citet{PAPER_limit_2015}, \citet{MWA_limit_2016_1}, \citet{MWA_limit_2016_2}, \citet{LOFAR_limit_2020}, and \citet{HERA_limit_2021}). \citet{Ung2020} discuss the non-trivial relationship between the mutual coupling of low noise amplifiers (LNAs) and receiver noise temperature, describing how such coupling effects can either increase or decrease receiver temperature depending on whether internal or external noise power dominates the receiver. Such coupling concerns must be carefully considered when comparing scientific results between different interferometric arrays.

Recently, \citet{Kern:2019} studied a subset of the possible antenna-antenna coupling we will consider. Their paper demonstrates a filtering and removal strategy which is resistant to EoR signal loss. However, the scope of their coupling model is limited to when the two antennae comprising a visibility reflect into one another. This does not consider the much broader set of coupling interactions (described in this paper) whereby a third antenna (e.g. Antenna k, $k\neq {i, j}$) can couple into a visibility term $V_{ij}$. The limitations of only studying the previous subset are more carefully described in Sections (\ref{sec:array_element_coupling}) and (\ref{sec:cross_features}) of this paper. 

The structure of this paper is as follows. In Section (\ref{sec:FORMALISM}), we derive a semi-analytic interferometric visibility equation which considers first-order antenna-antenna coupling. In Section (\ref{sec:ZERO_ORDER_SIM}), we discuss assumptions made in order to simulate zeroth-order visibilities in a HERA-like array. Through this paper, a 'zeroth-order' visibility refers to a standard interferometric visibility which does not consider antenna-antenna coupling. Section (\ref{sec:FIRST_ORDER_SIM}) applies the formalism of Section (\ref{sec:FORMALISM}), using the zeroth-order visibilities of Section (\ref{sec:ZERO_ORDER_SIM}) as input, to generate visibility data which is coupled to first order. We describe in Section (\ref{sec:PHENOMENON}) the phenomenology of features present in the first-order visibility data, which are not present in the zeroth order visibility data. In Section (\ref{sec:PSPEC}), we address how first-order antenna-antenna coupling features contaminate 21cm power spectrum measurements. Section (\ref{sec:DISCUSSIONS}) discusses limitations of our semi-analytic model and the challenges posed by first-order coupling in discriminating between the 21cm signal and astrophysical foregrounds. We also discuss how our model can, nevertheless, be used to design arrays and plan LST observation strategies which mitigate first-order coupling effects. Section (\ref{sec:CONCLUSION}) summarizes our findings and conclusions from throughout the paper. 

\section{General Formalism}
\label{sec:FORMALISM}
\begin{figure}
    \centering
    \includegraphics[width=.45\textwidth]{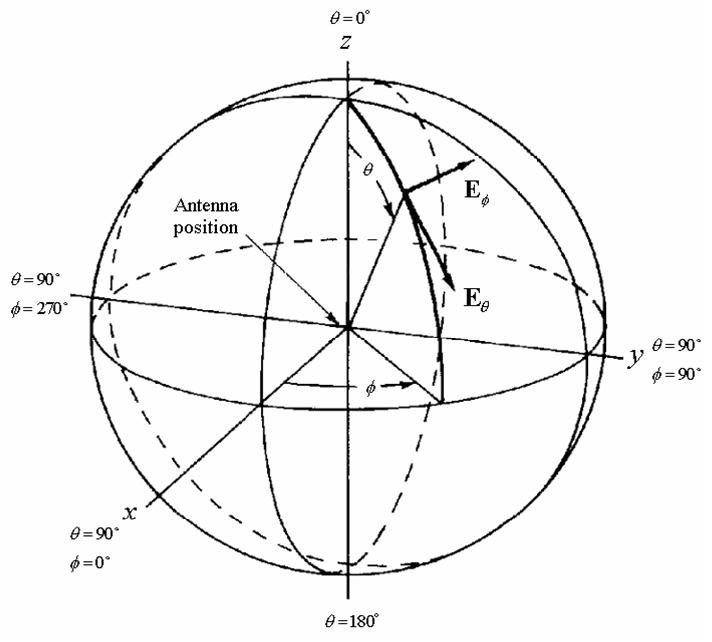}
    \caption{Spherical coordinate system, which we adopt as the basis to represent the components of our electric field terms. }
    \label{fig:Spherical_Coordinates}
\end{figure}
\subsection{Considering Array Element Coupling}
\label{sec:array_element_coupling} 
\begin{figure}
    \centering
    \includegraphics[width=.46\textwidth]{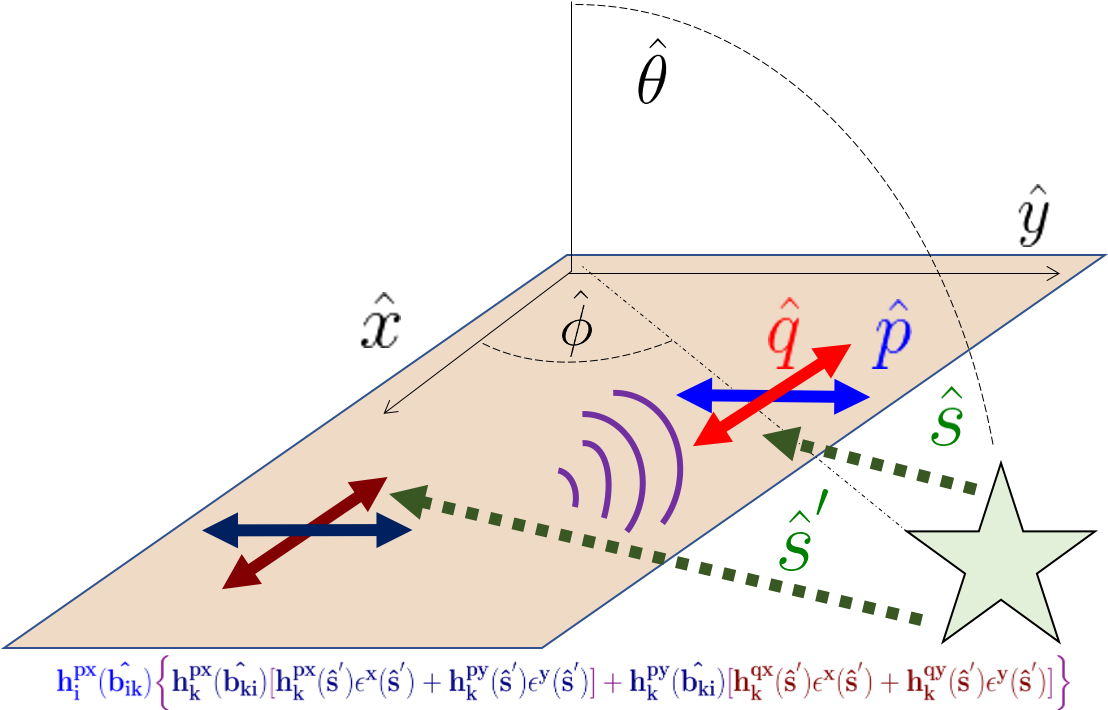}
    \caption{A conceptual outline of our coupling model, labeled by instrumental polarization and the coordinate basis vectors of Figure (\ref{fig:Spherical_Coordinates}). Incident astrophysical radiation with electric field basis vectors $\hat{\phi}$ and $\hat{\theta}$ hits the $i^{th}$ and $k^{th}$ elements in the array with electric field distributions $\epsilon(\hat{s})$ and $\epsilon(\hat{s}^{'})$, respectively. The $k^{th}$ element re-radiates some fraction of $\epsilon(\hat{s}^{'})$ to the $i^{th}$ element. Both elements have the same instrumental polarizations, $p$ (blue/dark blue) and $q$ (red/ dark red). The voltage induced by each polarization of antenna $i$ contains not only a response from  $\epsilon(\hat{s})$, but a contributions from the scattered electric field distribution from $k$. This scattered field contains components of \emph{both} polarizations of antenna $k$. The voltage response at the terminal of the $p^{th}$ pol of the $i^{th}$ interferometric element, caused by the $\hat{\phi}$ basis vector component of the electric field surface distribution which was re-radiated from the $k^{th}$ element is written below the figure, and is the first-order coupling component associated with equation (\ref{eq:v_i_1}). The scattered electric field distribution is depicted in purple because, when it arrives it either pol $p$ (blue) or pol $q$ (blue) of the $i^{th}$ element, it contains components of both instrumental polarizations of the $k^{th}$ element (dark red, dark blue). }
    \label{fig:coupling_pols_conceptual}
\end{figure}
We begin by writing the voltage responses of two antennae, as they would contribute to the standard (non-coupled) interferometric visibility equation. For completeness, we include a derivation of this specific form of the voltage response equations in Appendix \ref{sec:general_formalism_motivation}. Let the $p^{th}$ pol of the $i^{th}$ interferometric element be characterised by an effective height of $h^{p\phi}_{i}(\hat{s})$, where $\hat{\phi}$ is a basis vector component of an incident electric field. The voltage response ($v^{p\phi}_{i}$) at the terminal of the $p^{th}$ pol of this $i^{th}$ element, caused by the $\hat{\phi}$ component of the incident electric field surface distribution ($\epsilon^\phi(\hat{s})$) is

\begin{equation}\label{eq:v_i_0}
v^{p\phi}_{i} = \int h^{p\phi}_{i}(\hat{s})  \varepsilon^{\phi}(\hat{s})\frac{e^{2\pi i \frac{\nu}{c}|R\hat{s}-\vec{b_i}| }}{|R\hat{s}-\vec{b_i}|}R^2 d\Omega
\end{equation}

Similarly, the voltage response at the terminal of the $q^{th}$ pol of the $j^{th}$ interferometric element, caused by the surface distribution of the $\hat{\theta}$ basis vector component of the electric field is
\begin{equation}\label{eq:v_j_0}
v^{q\theta}_{j} = \int h^{q\theta}_{j}(\hat{s}^{``}) \varepsilon^{\theta}(\hat{s}^{``})\frac{e^{2\pi i \frac{\nu}{c}|R\hat{s}^{``}-\vec{b_j}| }}{|R\hat{s}^{``}-\vec{b_j}|}R^2 d\Omega^{``}
\end{equation}

The above formalism would be complete if there were only two interferometric elements in the array and each element perfectly absorbed all incident radiation. We wish to explore the realistic situation, however, where there are greater than two elements in the array and none of the elements are perfectly impedance matched to their feed terminals. In this case, each element re-radiates some portion of the incident astrophysical electric field. Thus, each polarization of the interferometric elements will not only have a voltage induced by the incident astrophysical electric field, but have a voltage induced by re-radiation from all other elements in the array.

In this work, we only consider 1st-order re-radiation of the astrophysical electric field, which arrives at the $i^{th}$ and $j^{th}$ elements after being previously re-radiated by only one other element $k$, not more than one element. Figure (\ref{fig:coupling_pols_conceptual}) conceptually describes the re-radiation model from antenna $i$ to antenna $k$. Because the array may be located generally on the geographic (2D) plane, not simply in a (1D) line, the scattered electric field (from the $k^{th}$ element), upon arrival to any instrumental polarization of all other elements, can be a function of $\emph{both}$ instrumental polarizations of the radiating ($k^{th}$) element. Consequently, the scattered electric field can be a function of both incident astrophysical electric field basis vectors. We do not yet characterize the scattered electric field; this will be done in Section (\ref{sec:scattered_field_epsillon}). For now, we generally define the scattered field distribution received by the $i^{th}$ element in terms such as $\varepsilon^{s\phi,p\theta}(\hat{s}, \hat{b_{ki}})$. The first superscript of the electric field (e.g. '$\phi$') refers to $\hat{\phi}$ component of what is received by the $i^{th}$ element from the radiating $k^{th}$ element. After the comma in the superscript (e.g. $p\theta$), we denote the incident electric field basis vector (e.g. $\hat{\theta}$) contribution to each instrument polarization (e.g. $p$) of the radiating $k^{th}$ element. A similar, general expression is used for the electric field distribution which is scattered from the $k^{th}$ element to the $j^{th}$ element, e.g. $\varepsilon^{s\theta,q\phi}(\hat{s}, \hat{b_{kj}})$

To summarize, the voltage response at the terminal of the $p^{th}$ pol of the $i^{th}$ interferometric element, caused by the $\hat{\phi}$ basis vector component of the electric field surface distribution which was re-radiated from the $k^{th}$ element may contain components from both instrumental polarizations and both incident electric field basis vectors of the $k^{th}$ element. Similarly, the voltage response at the terminal of the $q^{th}$ pol of the $j^{th}$ interferometric element, caused by the $\hat{\theta}$ basis vector component of the electric field surface distribution which was re-radiated from the $k^{th}$ element may contain components from both instrumental polarizations and both incident electric field basis vectors of the $k^{th}$ element. We formalize this model of antenna-to-antenna coupling in Appendix \ref{sec:general_formalism_motivation} by replacing equations (\ref{eq:v_i_0}) and (\ref{eq:v_j_0}) with the voltage responses presented in equations (\ref{eq:v_i_1}) and (\ref{eq:v_j_1}), respectively.

The raw data product of our interferometer is the correlation of the voltage responses from any two elements in the array, $\langle v_i v^*_j\rangle$, which is referred to as a "visibility", $V_{ij}$. If the two elements are the same ($i = j$), the visibility is the product of an autocorrelation. If the two elements are not the same ($i \neq j$), the visibility is the product of a cross-correlation. In either case, the correlation data product is useful because it exploits the spatial coherence of the astrophysical electric field components at two different test points, $\langle E^\phi(\hat{s}) E^{\theta*}(\hat{s}^{``})\rangle$. In the next section, we will use equations (\ref{eq:v_i_1}) and (\ref{eq:v_j_1}) to calculate the visibility, accounting for first-order antenna-antenna coupling.

\subsection{General First-Order Visibility Formalism}
\label{sec:general_matrix_formalism} 
The voltage responses presented in equations (\ref{eq:v_i_1}) and (\ref{eq:v_j_1}) are expressed as functions of feed polarization ($p, q$), interferometric elements ($i, j$), and electric field basis vector components ($\phi, \theta$) which are tangent to the celestial sphere. These expressions can be re-written in a matrix formalism, to represent all polarizations and basis vector components of the visibility $\textbf{V}^{0,1}_{ij}$, whether to zeroth order in coupling (no-coupling) or including first order antenna-to-antenna scattering. Let us define voltage vectors $\vec{v_i}$ and $\vec{v_j}$ to characterize, per interferometric element, the induced voltage at the terminal of both feed polarizations.
\begin{equation}\label{eq:voltage_vectors}
\vec{v_i}=
\begin{bmatrix}
v^p_i\\
v^q_i
\end{bmatrix},
 \vec{v_j}=
\begin{bmatrix}
v^p_j\\
v^q_j
\end{bmatrix}
\end{equation} 
Letting "$\dagger$" represents the conjugate-transpose, the visibility matrix may be defined as
\begin{align}\label{eq:V_matrix_define}
\textbf{V}^{0,1}_{ij}(\nu) & =
\begin{bmatrix}
V^{pp}_{ij} & V^{pq}_{ij}\\
V^{qp}_{ij} & V^{qq}_{ij}
\end{bmatrix}^{0,1}
=
\begin{bmatrix}
\langle v^p_i v^{*p}_j\rangle & \langle v^p_i v^{*q}_j\rangle\\
\langle v^q_i v^{*p}_j\rangle & \langle v^q_i v^{*q}_j\rangle
\end{bmatrix}^{0,1}
\\ & = \langle \vec{v_i}\times\vec{v}^{\dagger}_j\rangle^{0,1} \nonumber
\end{align}
Similarly, components of electric field surface distributions (for both incident astrophysical radiation and re-radiation for first-order coupling) can be represented as vectors, 

\begin{equation}\label{eq:epsilon_vectors}
\vec{\epsilon}=
\begin{bmatrix}
\epsilon_\phi\\
\epsilon_\theta
\end{bmatrix},
 \vec{\epsilon}^s=
\begin{bmatrix}
\epsilon^s_\phi\\
\epsilon^s_\theta
\end{bmatrix}
\end{equation} 

Since the effective height, $h^\phi_p(\hat{s},\nu)$ is written in terms of feed polarization and electric field basis vectors, it, too, can easily be re-written in matrix formalism. Following the unit convention in experimental astrophysics (which may differ from optics, or other sub-disciplines), we define a "Jones Matrix", $\textbf{J}$ with units of length to represent the effective height, where
\begin{equation}\label{eq:Jones_Matrix_Relationship}
v=
\begin{bmatrix}
v_p\\
v_q
\end{bmatrix}=\textbf{J}\vec{\epsilon}=
\begin{bmatrix}
h^\phi_p & h^\theta_p\\
h^\phi_q & h^\theta_q
\end{bmatrix}
\begin{bmatrix}
\epsilon_\phi\\
\epsilon_\theta
\end{bmatrix}
\end{equation}
Using equations (\ref{eq:v_i_1}) and (\ref{eq:v_j_1}) for $\vec{v_i}$ and $\vec{v_j}$, respectively, We calculate the first-order visibility terms, $\langle \vec{v_i}\times\vec{v}^{\dagger}_j\rangle^{1}$,  in matrix formalism.

    \begin{align}\label{eq:V_ij_1}
    \mathbf{V}^{1}_{ij} & =\int \int \bigg(  \langle\mathbf{J_i}(\hat{s})\vec{\epsilon}(\hat{s}) \times(\mathbf{J_j}(\hat{s}^{``})\vec{\epsilon}(\hat{s}^{``}))^\dagger\rangle \\ 
    & \hspace{3.1cm} \frac{e^{2\pi i \frac{\nu}{c}(R\hat{s}-\vec{b_i}-R\hat{s}^{``}+\vec{b_j})}}{|R\hat{s}-\vec{b_i}||R\hat{s}^{``}-\vec{b_j}|}R^4 d\Omega d\Omega^{``} \bigg) \nonumber \\
    & + \sum_{k \neq i}\int \int \bigg(  \langle\mathbf{J_i}(\hat{b_{ik}})\vec{\epsilon}^{s}_{ki}(\hat{s}^{'}, \hat{b_{ki}}) \times(\mathbf{J_j}(\hat{s}^{``})\vec{\epsilon}(\hat{s}^{``}))^\dagger\rangle \nonumber \\
    & \hspace{3.1cm} \frac{e^{2\pi i \frac{\nu}{c}(R\hat{s}^{'}-\vec{b_k}-R\hat{s}^{``}+\vec{b_j})}}{|R\hat{s}^{'}-\vec{b_k}||R\hat{s}^{``}-\vec{b_j}|}R^4 d\Omega^{'} d\Omega^{``} \bigg) \nonumber \\
    & + \sum_{k \neq j}
    \int \int \bigg( \langle\mathbf{J_i}(\hat{s})\vec{\epsilon}(\hat{s}) \times(\mathbf{J_j}(\hat{b_{jk}})\vec{\epsilon}^{s}_{kj}(\hat{s}^{'}, \hat{b_{kj}}))^\dagger\rangle \nonumber \\
    & \hspace{3.1cm} \frac{e^{2\pi i \frac{\nu}{c}(R\hat{s}-\vec{b_i}-R\hat{s}^{'}+\vec{b_k})}}{|R\hat{s}-\vec{b_i}||R\hat{s}^{'}-\vec{b_k}|}R^4 d\Omega d\Omega^{'} \bigg) \nonumber
    \end{align}


Note that equations (12) and (13) from \citet{Kern:2019} are special cases of our equation (\ref{eq:V_ij_1}) which only include the reflection from antenna $j$ to $i$ (not other elements, $k$). The \cite{Kern:2019} coupling model produces auto-correlation coupling terms in visibility, since $b_{jj}=0$, but does not consider cross-correlation terms which are present in (\ref{eq:V_ij_1}). By Assumption A (stated clearly in Appendix \ref{sec:general_formalism_motivation}, we can simplify equation (\ref{eq:V_ij_1}), for $|R\hat{s}-\vec{b_i}| \approx R - \vec{b_i}\cdot\hat{s}$. This removes all terms of $R$ from the exponential factors in (\ref{eq:V_ij_1}). The denominators in the same equation reduce further. For example, $|R\hat{s}-\vec{b_i}||R\hat{s}^{``}-\vec{b_j}| \approx R^2$, regardless of if we are integrating in the differential direction $\hat{s}$ or $\hat{s}^{``}$.

Finally, we note that the expectation values of each term of (\ref{eq:V_ij_1}) may be re-written solely in terms of the expectation value of the electric field surface distributions. For example, in the $\sum_{k \neq i}$ term:

\begin{align}\label{eq:expectation_E_only}
\langle\mathbf{J_i}(\hat{b_{ik}})\vec{\epsilon}^{s}_{ki}(\hat{s}^{'}, \hat{b_{ki}}) \times & (\mathbf{J_j}(\hat{s}^{``})\vec{\epsilon}(\hat{s}^{``}))^\dagger\rangle = \\ & \mathbf{J_i}(\hat{b_{ik}})\langle\vec{\epsilon}^{s}_{ki}(\hat{s}^{'}, \hat{b_{ki}})\times\vec{\epsilon}(\hat{s}^{``})^\dagger\rangle\mathbf{J_j}(\hat{s}^{``})^\dagger \nonumber
\end{align}
This latter form will be useful for simplifying equation (\ref{eq:V_ij_1}) in the forthcoming two sections.

\subsection{Characterizing the Scattered Electric Field}\label{sec:scattered_field_epsillon}
To simplify (\ref{eq:V_ij_1}), we start by examining the scattered electric field terms associated with re-radiation of the incident astrophysical radiation. Standard antenna scattering formalisms, such as \citet{FLOKAS}, characterize the total electric field scattered from the  $p^{th}$ feed polarization of a single $k^{th}$ element in a large array due to impedance mismatch as
\begin{align}\label{eq:E_scatter}
    \vec{E}_{ki}^{s,p}( \hat{s}, \hat{b_{ki}}) &= \frac{i\eta_{0}}{4\lambda R_{k}} \vec{h}^{p}_{k}(\hat{b_{ki}})(\vec{h}^{p}_{k}(\hat{s})\cdot\vec{E}(\hat{s}))\frac{ e^{-2 \pi i \frac{\nu}{c} | \vec{b_{ki}}| }}{| \vec{b_{ki}}|}\Gamma_{k}
\end{align}
where $\Gamma_{k}(\nu)$ is the frequency-dependent reflection coefficient of the $k^{th}$ feed, $\eta_{0}$ is the impedance of free space, and $R_{k}$ is the real part of the impedance of the $k^{th}$ antenna. This expression assumes the scattered electric field may be approximated by the far-field limit when it is received by another interferometric element. 

The formalism of (\ref{eq:E_scatter}) assumes a spherical wave propagation convention of $e^{-2\pi i \frac{\nu}{c} |\vec{b_{ki}}|}$. To make this formalism consistent with our chosen convention (See equation \ref{eq:Propagation} in Appendix \ref{sec:general_formalism_motivation}), we flip the direction of the complex exponential and integrate over the appropriate electric field surface distribution (\ref{eq:E_at_i}), which involves an integral of the general form $\int e^{2\pi i \frac{\nu}{c} s}ds = \frac{-i}{2\pi}e^{2\pi i \frac{\nu}{c} s} + c$, for some arbitrary constant $c$. Using a matrix formalism to accounting for all instrumental polarizations,
\begin{align}\label{eq:E_scatter_correct_propagator}
    \vec{\epsilon}_{ki}^{s,p}( \hat{s}, \hat{b_{ki}}) &= \frac{-i\eta_{0}}{4\lambda R_{k}} \mathbf{J}_{k}(\hat{b_{ki}})(\mathbf{J}_{k}(\hat{s})\cdot\vec{\epsilon}(\hat{s}))\frac{ e^{2 \pi i \frac{\nu}{c} | \vec{b_{ki}}| }}{| \vec{b_{ki}}|}\Gamma_{k}
\end{align}

It is important to note that our formalism does not consider all re-radiation reflections which have occurred in the array since the initial time that an incident wavefront reached the $i^{th}$ (eq. \ref{eq:v_i_1}) or $j^{th}$ (eq. \ref{eq:v_j_1}) antennae, respectively. Equation (\ref{eq:E_scatter_correct_propagator}) only considers how an incident wavefront gets re-radiated by each element one time. For our formalism to meaningfully describe the scattering in the array, we make a third assumption, Assumption C, that the incident astrophysical radiation changes at time scales much longer than the timescale of reflections in the array (or at least those which have an amplitude that could meaningfully induce a terminal voltage above the noise floor of our analog chain). For most purposes, we consider the sky to not have 'moved' over the array during a visibility measurement integrated over the order of minutes, and reflections in the array to have occurred up to thousands of nanoseconds. Thus, the steady state assumptions of our formalism are reasonable, and gives us the opportunity to, in a closed and semi-analytic form, approximate first-order antenna-to-antenna coupling. 

Inserting (\ref{eq:E_scatter_correct_propagator}) into the expectation values of (\ref{eq:V_ij_1}) as rewritten in the form of (\ref{eq:expectation_E_only}), we notice the per-basis-vector expectation quantities in both coupling terms, namely $\langle\varepsilon^{s,p\phi}(\hat{s^{'}}, \hat{b_{ki}}) \varepsilon^{*\theta}(\hat{s}^{``})\rangle$ and $\langle\varepsilon^{\phi}(\hat{s}) \varepsilon^{*s,q\theta}(\hat{s}^{'}, \hat{b_{kj}})\rangle$, can be rewritten in terms of effective heights multiplied by the same expectation found in the zeroth-order term of (\ref{eq:V_ij_1}), e.g. $\langle\varepsilon^{\phi}(\hat{s}) \varepsilon^{*\theta}(\hat{s}^{``})\rangle$. For example, in the $\sum_{k \neq i}$ term,

\begin{align}\label{eq:expectation_E_and_scattered_E}
& \mathbf{J_i}(\hat{b_{ik}})\langle\vec{\epsilon}^{s}_{ki}(\hat{s}^{'}, \hat{b_{ki}})\times \vec{\epsilon}(\hat{s}^{``})^\dagger\rangle\mathbf{J_j}(\hat{s}^{``})^\dagger = \\
& \frac{-i\eta_{0}}{4\lambda R_{k}}\frac{ e^{2 \pi i \frac{\nu}{c} | \vec{b_{ki}}| }}{| \vec{b_{ki}}|}\Gamma_{k}\mathbf{J_i}(\hat{b_{ik}})\mathbf{J_k}(\hat{b_{ki}})\mathbf{J_k}(\hat{s^{'}})\langle\varepsilon^{\phi}(\hat{s^{'}}) \varepsilon^{*\theta}(\hat{s}^{``})\rangle \mathbf{J_j}(\hat{s}^{``})^\dagger \nonumber 
\end{align}

\subsection{Full Stokes, First-Order Visibility Formalism, with Summary of Key Assumptions}\label{sec:first_order_visibility_and_assumptions}
We may simplify the zeroth order terms in the form of (\ref{eq:expectation_E_and_scattered_E}) by making a fourth assumption, Assumption D, that the radiation from astrophysical sources is not spatially coherent. Standard derivations of visibility equations also make this assumption (e.g. \citet{1999ASPC..180....1C}), which relies on the Van Cittert–Zernike theorem. Physically, we may interpret Assumption D as saying that the electromagnetic radiation from distant astrophysical sources, at the frequencies over which we calculate visibilities, is the result of random, thermal processes, such as synchrotron radiation. Mathematically, Assumption D asserts that $\langle\varepsilon^{\phi}(\hat{s}) \varepsilon^{*\theta}(\hat{s}^{'})\rangle = 0$, for $\hat{s} \neq \hat{s^{'}}$.

There are three noteworthy simplifications in (\ref{eq:V_ij_1}) due to Assumption D. First, the effective delay in the exponential terms is based only on the distance between baselines, for example $\vec{b_j} - \vec{b_i} = \vec{b_{ij}}$. Second, there is no more radial dependence on $\vec{R}$ in (\ref{eq:V_ij_1}). Third, expectation values may be re-written such that we only need to integrate over one solid angle, not two. 
\begin{equation}\label{eq:expectation_to_coherency}
\langle\varepsilon^{\phi}(\hat{s}) \varepsilon^{*\theta}(\hat{s}^{'})\rangle = \frac{C^{\phi\theta}(\hat{s})\delta_{D} (\hat{s}-\hat{s^{'}})}{|\vec{R}|^2}
\end{equation}

The spatial coherence of the electric field may be represented as a "coherency matrix" with the same relationship to electric field surface distributions found in (\ref{eq:expectation_to_coherency}).
\begin{equation}\label{eq:expectation_to_coherency_matrix}
\langle\varepsilon(\hat{s})\times\varepsilon^{\dagger}(\hat{s}^{'})\rangle=\frac{\textbf{C}(\hat{s})\delta_{D}(\hat{s}-\hat{s}^{'})}{|\vec{R}|^2}
\end{equation}

\begin{equation}\label{eq:C_matrix}
\textbf{C}(\hat{s})=
\begin{bmatrix}
\langle \epsilon_{\phi}(\hat{s}) \epsilon^{*}_{\phi}(\hat{s})\rangle & \langle \epsilon_{\phi}(\hat{s}) \epsilon^{*}_{\theta}(\hat{s})\rangle\\
\langle \epsilon_{\theta}(\hat{s}) \epsilon^{*}_{\phi}(\hat{s})\rangle & \langle \epsilon_{\theta}(\hat{s}) \epsilon^{*}_{\theta}(\hat{s})\rangle
\end{bmatrix}
\end{equation}

Components of the coherency matrix (\ref{eq:C_matrix}) are represented in terms of Stokes parameters in various references in the literature. In \citet{THOMPSON},
$\langle|\epsilon_\phi|^2\rangle=\frac{I+Q}{2}$ and $\langle|\epsilon_\theta|^2\rangle=\frac{I-Q}{2}$.  A full-Stokes representation of the coherency matrix is discussed in \citet{Smirnov_2011},
\begin{equation}\label{eq:C_matrix_Stokes}
\textbf{C}(\hat{s})=
\begin{bmatrix}
I+Q & U+iV\\
U-iV & I-Q
\end{bmatrix}
\end{equation}
Representing the phase difference between $\epsilon_\phi$ and $\epsilon_\theta$ as $\delta$, Smirnov's formalism is consistent with Thompson and Swenson's when

\begin{equation}\label{eq:STOKES} 
    \begin{alignedat}{2}
    I=\frac{1}{2}(\langle|\epsilon_\phi|^2\rangle + \langle|\epsilon_\theta|^2\rangle) 
    &\hspace{0.5cm} Q=\frac{1}{2}(\langle|\epsilon_\phi|^2\rangle - \langle|\epsilon_\theta|^2\rangle) 
    \\
    U=\langle|\epsilon_\phi||\epsilon_\theta|cos\delta\rangle 
    &\hspace{0.5cm} V=\langle|\epsilon_\phi||\epsilon_\theta|sin\delta\rangle 
    \end{alignedat}
\end{equation}

Inserting (\ref{eq:E_scatter_correct_propagator}) and (\ref{eq:expectation_to_coherency}) into (\ref{eq:V_ij_1}), and recalling the first two simplifications resulting from Assumption D, (\ref{eq:V_ij_1}) may be re-written as
    \begin{align}\label{eq:V_ij_1_post_C}
    \mathbf{V}^{1}_{ij} & = \int \mathbf{J}_i(\hat{s})\langle\vec{\epsilon}(\hat{s})\times\vec{\epsilon}^\dagger (\hat{s})\rangle\mathbf{J}_j^\dagger(\hat{s}) e^{2\pi i \frac{\nu}{c}\vec{b_{ij}}\cdot\hat{s}}d\Omega \\
    & - \sum_{k \neq i}\bigg\{ \frac{i \eta_0 \Gamma_k}{4\lambda R_k |\vec{b_{ki}}|} e^{2\pi i \frac{\nu}{c}|\vec{b}_{ki}|}\mathbf{J}_i(\hat{b_{ik}})\mathbf{J}_k(\hat{b_{ki}}) \nonumber \\
    & \hspace{1.5cm} \bigg( \int \mathbf{J}_k(\hat{s})\langle\vec{\epsilon}(\hat{s})\times\vec{\epsilon}^\dagger (\hat{s})\rangle\mathbf{J}_j^{\dagger}(\hat{s}) e^{2\pi i \frac{\nu}{c}\vec{b_{kj}}\cdot\hat{s}}d\Omega \bigg) \bigg\} \nonumber \\
    & + \sum_{k \neq j} \bigg\{ \bigg( \int \mathbf{J}_{i}(\hat{s})\langle \vec{\epsilon}(\hat{s})\times\vec{\epsilon}^\dagger (\hat{s})\rangle\mathbf{J}^{\dagger}_{k}(\hat{s}) e^{2\pi i \frac{\nu}{c}\vec{b_{ik}}\cdot\hat{s}}d\Omega \bigg) \nonumber \\
    & \hspace{2.3cm} \frac{i \eta_0 \Gamma^*_k}{4\lambda R_k |\vec{b_{kj}}|} e^{-2\pi i \frac{\nu}{c}|\vec{b}_{kj}|}\mathbf{J}_{k}(\hat{b_{kj}})\mathbf{J}_{j}(\hat{b_{jk}}) \bigg\} \nonumber
    \end{align}
If antenna-to-antenna coupling had not been considered, the visibility equation would only include the first of the three terms in (\ref{eq:V_ij_1_post_C}). We can re-write this no-coupling term as 
\begin{align}\label{eq:V_ij_0_matrix}
\mathbf{V}^0_{ij} & = \int \mathbf{J}_i(\hat{s})\langle\vec{\epsilon}(\hat{s})\times\vec{\epsilon}^\dagger (\hat{s})\rangle\mathbf{J}_j^\dagger(\hat{s})e^{2\pi i \frac{\nu}{c}\vec{b_{ij}}\cdot\hat{s}}d\Omega \\
& = \int \textbf{J}_i(\hat{s})\textbf{C}(\hat{s})\textbf{J}^{\dagger}_j(\hat{s})e^{2\pi i \frac{\nu}{c}\vec{b_{ij}}\cdot\hat{s}}d\Omega \nonumber
\end{align}

Notice that the second and third ("coupling") terms of (\ref{eq:V_ij_1_post_C}) depend on zeroth order ("no-coupling") visibilities of various baselines $V^0_{kj}$ and $V^0_{ik}$.  Re-writing (\ref{eq:V_ij_1_post_C}) in terms of zeroth-order visibilities,

\begin{align}\label{eq:V_ij_1_with_0_matrix}
    \textbf{V}^1_{ij} & = \textbf{V}^0_{ij} \\
    & + \frac{i \eta_0}{4\lambda}\bigg(-\sum_{k \neq i}\frac{\Gamma_k}{R_k |\vec{b_{ki}}|} e^{2\pi i \frac{\nu}{c}|\vec{b}_{ki}|}\textbf{J}_{i}(\hat{b_{ik}})\textbf{J}_{k}(\hat{b_{ki}}) \textbf{V}^0_{kj} \nonumber \\
    & + \hspace{1cm} \sum_{k \neq j}\frac{\Gamma^*_k}{ R_k |\vec{b_{kj}}|} e^{-2\pi i \frac{\nu}{c}|\vec{b}_{kj}|} \textbf{V}^0_{ik}\textbf{J}^{\dagger}_{k}(\hat{b_{kj}})\textbf{J}^{\dagger}_{j}(\hat{b_{jk}}) \bigg) \nonumber
\end{align}

Note that all terms associated with re-radiated (mutually-coupled) power introduce copies of zeroth-order visibilities. These copies are attenuated by extra beam factors ($\mathbf{JJ}$ or $\mathbf{J}^\dagger \mathbf{J}^\dagger$), decayed in magnitude by the baseline length, phase-shifted,and then delayed by the time it takes for the reflected radiation to travel the coupled baseline vector. Lastly, recall that the derivation of equation (\ref{eq:V_ij_1_with_0_matrix}) requires four assumptions, which we previously defined as Assumptions A-D. Assumptions A-B are written in Appendix \ref{sec:general_formalism_motivation}. Section (\ref{sec:scattered_field_epsillon}) notes Assumption C, and Assumption D was noted in this section. All four assumptions are summarized in the table below. 

\begin{tabular}{ |p{1.8cm}||p{5.4cm}| }
 \hline
 \multicolumn{2}{|c|}{Key Assumptions} \\
 \hline
 Assumption & Description\\
 \hline
 A & $|\vec{R}| >> |\vec{b}_i|$ \\
 B & Free space (vacuum) propagation \\
 C & $\Delta t$ (sky change) $>> \tau$ (delays of scattering) \\
 D & Van Cittert–Zernike theorem \\
 \hline
\end{tabular}

\section{Simulation of Zeroth Order Visibilities}\label{sec:ZERO_ORDER_SIM}
Armed with a general formalism for first-order array element coupling, let us establish a simulated framework for studying its effects. Specifically, we simulate zeroth-order and first-order visibility data for HERA using a simplified version of the formalism. HERA is a 350-element interferometer, built in South Africa’s Karoo Radio Astronomy Reserve. HERA was designed specifically to characterize the evolution of the 21cm signal from cosmic dawn (z$\approx$30) through the full reionization of the IGM (z$\approx$6). Each element of the interferometer is a 14m parabolic dish, capable of observing from 50-250 MHz. For this analysis, we limit our simulations to the frequency range 144-169 MHz. Active electronic receiver components are currently being installed onto HERA's Vivaldi-style dish feeds; both the receiver system and the feed are characterised in \cite{Fagnoni2020}. 

\begin{figure}
    \centering
    \includegraphics[width=.42\textwidth]{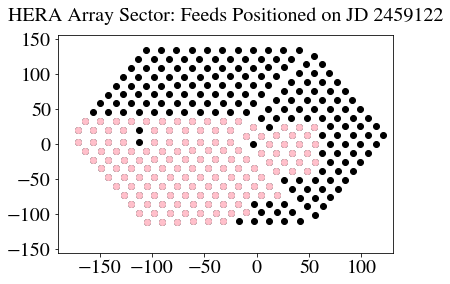}
    \caption{A 143 element subset (pink) of the HERA array, which contains a total of 350 elements. The X (east) and Y (north) axes have units of meters, in arbitrary, local ENU coordinates. This array subset corresponds to all Vivaldi feeds which were positioned on Julian date 2459122, a night during which a commissioning observation was recorded by the HERA collaboration.}
    \label{fig:Array_Plot}
\end{figure}

While it is beyond the scope of this work to directly compare the results of our simulations to observed HERA data, we nonetheless choose to study the HERA array as it was configured for a commissioning observation on Julian date (JD) 2459122, during which the HERA collaboration recorded data. Only 143 of the 350 Vivaldi feeds were positioned and installed on this JD. Simulating this 143-element subset, rather than the full array, will make it convenient to apply the results of this work to a follow-up analysis which directly compares first-order interferometric visibilities to HERA data. Figure (\ref{fig:Array_Plot}) plots the 143-element array configuration as a function of antenna ENU coordinates.

In preparation for the analysis of our simulated HERA visibilities, we confirm that the far-field approximation used in Equation (\ref{eq:E_scatter_correct_propagator}) is valid across the HERA array. We also analyze simulated HERA beam products and derive how these products, which were calculated using standard electromagnetic simulations software that implicitly accounts for antenna impedance, can be unit-wise compatible with our general formalism, which is explicitly dependent on antenna impedance. See Appendix (\ref{sec:BEAM_COMPATIBILITY}) for our work to reach these conclusions. The HERA beam products were simulated using CST Microwave Studio \footnote{https://www.3ds.com/products-services/simulia/products/cst-studio-suite/}. \citet{Fagnoni2020} describes in detail the electromagnetic properties of the simulated HERA beam.  

\subsection{A Simplified Formalism to Simulate}
\label{sec:sim_simple}
Let us now make two simplifications to the general formalism in order to reduce the computational expense of the simulation over a HERA-like array. First, we only consider an unpolarized astrophysical sky. For now, this consists of the Stokes I component of a combined diffuse (GDSM) and point source (GLEAM) sky. In this case, the coherency matrix greatly simplifies, since the polarized Stokes parameters $Q=U=V=0$. Second, we simulate all beams in the array as having identical electromagnetic properties. Thus, $\textbf{J}_i(\hat{s}) = \textbf{J}_j(\hat{s}) = \textbf{J}(\hat{s})$ . Further, $\Gamma_k = \Gamma_i = \Gamma$ and $R_k = R_i = R_{ant}$. With these simplifications, the first-order visibilities of (\ref{eq:V_ij_1_with_0_matrix}) simplify to
\begin{align}\label{eq:V_ij_1_with_0_simp}
    \textbf{V}^1_{ij} & = \textbf{V}^0_{ij} \\
    & + \frac{i \eta_0}{4\lambda}\bigg(-\sum_{k \neq i}\frac{\Gamma}{R_{ant} |\vec{b_{ki}}|} e^{2\pi i \frac{\nu}{c}|\vec{b}_{ki}|}\textbf{J}(\hat{b_{ik}})\textbf{J}(\hat{b_{ki}}) \textbf{V}^0_{kj} \nonumber \\
    & + \hspace{1cm} \sum_{k \neq j}\frac{\Gamma^*}{ R_{ant} |\vec{b_{kj}}|} e^{-2\pi i \frac{\nu}{c}|\vec{b}_{kj}|} \textbf{V}^0_{ik}\textbf{J}^{\dagger}(\hat{b_{kj}})\textbf{J}^{\dagger}(\hat{b_{jk}}) \bigg) \nonumber
\end{align}
where the zeroth order visibilities $\textbf{V}^0_{ij}$, formerly expressed as (\ref{eq:V_ij_0_matrix}) now simplify to
\begin{equation}\label{eq:V_ij_0_matrix_simp}
\textbf{V}^0_{ij} = \int I(\hat{s},
\nu) \textbf{J}(\hat{s})\textbf{J}^{\dagger}(\hat{s})e^{2\pi i \frac{\nu}{c}\vec{b_{ij}}\cdot\hat{s}}d\Omega
\end{equation}

\subsection{Zeroth Order Visibilities in a Simulated HERA-Like Array}
\label{sec:sim_0}
All zeroth order visibilities presented in this paper are simulated without noise, in the frequency range 144-169 MHz using the formalism of (\ref{eq:V_ij_0_matrix_simp}), as implemented by the Healvis visibility simulator presented in \citet{2019ascl.soft07002L}. Two sky models are simulated separately. First, a "diffuse" sky model derived from the source-subtracted and de-striped Haslam 408 MHz map presented in \citet{2015MNRAS.451.4311R}. Second, a point source model which includes the GLEAM I catalog along with point source models of 3C161, 3C409, and Cassiopeia A. Multi-component extended source models for Centaurus A, Hera A, Hydra A, Pictor A, and Virgo A, are also included, as well as an extended source model for Fornax A. For the purposes of this work, the second sky model is called 'GLEAM I + Extended Sources'; this model is described in detail in \citet{2021arXiv210711487B}. The zeroth-order visibilities of the two sky models are added together (in the complex plane) to generate a third set of zeroth-order visibility data. This third set shall be dubbed the 'Combined Sky' model, as it contains both diffuse structure and point sources.

\begin{figure}
    \centering
    \includegraphics[width=.47\textwidth]{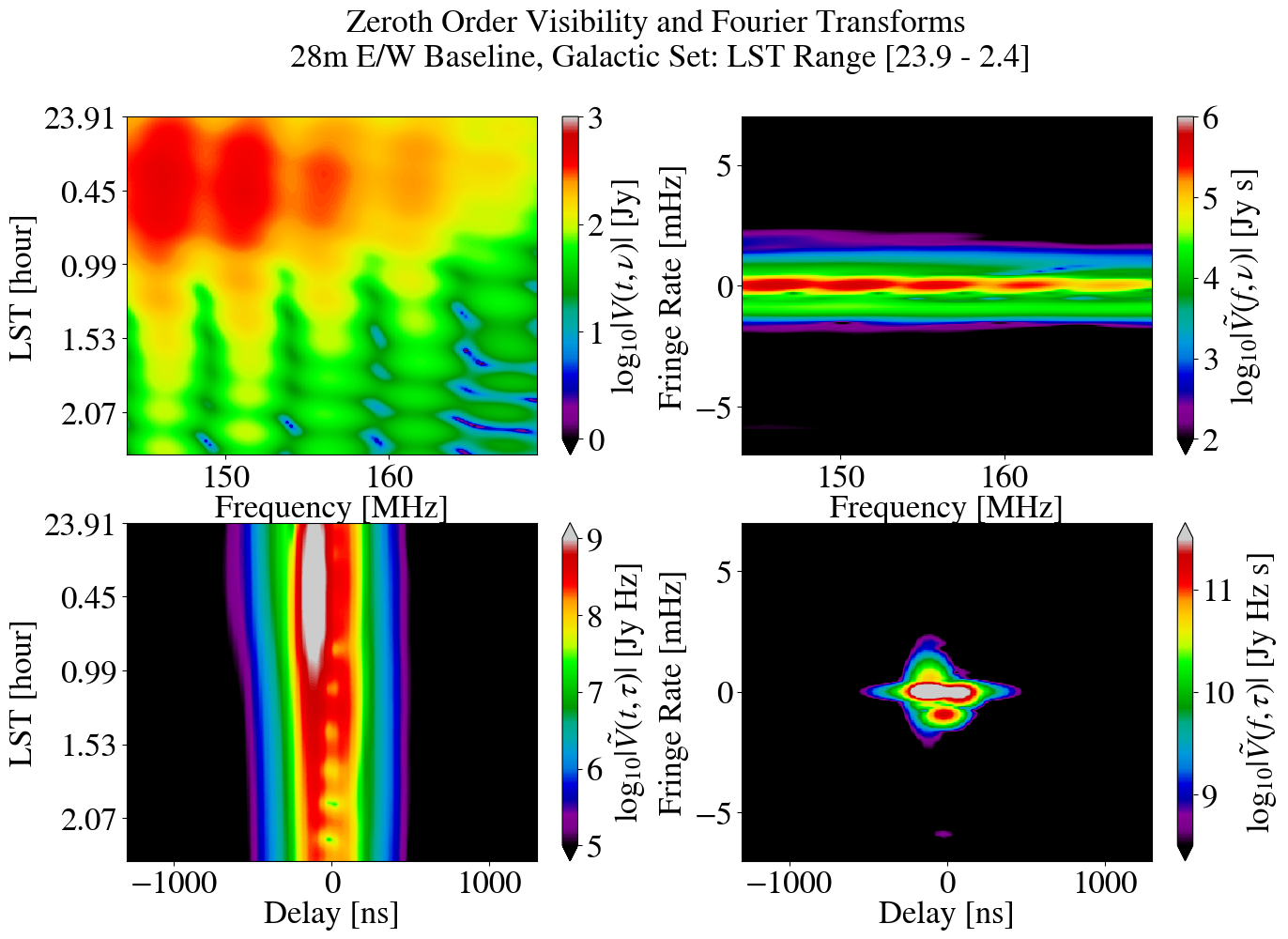}
    \caption{Zeroth-order visibility data of the combined (diffuse + point source) sky model,for a 28m E/W baseline, generated by the healvis simulator. Frequency range: 144-169 MHz. LST range: [23.9-2.4]. Moving clockwise from the top-left: time versus frequency ($t$ vs. $\nu$), fringe-rate versus frequency ($f$ vs. $\nu$), fringe-rate versus delay ($f$ vs. $\tau$), and time versus delay ($t$ vs. $\tau$).}
    \label{fig:zeroth_28m_EW_combined_GS}
\end{figure}

\begin{figure}
    \centering
    \includegraphics[width=.47\textwidth]{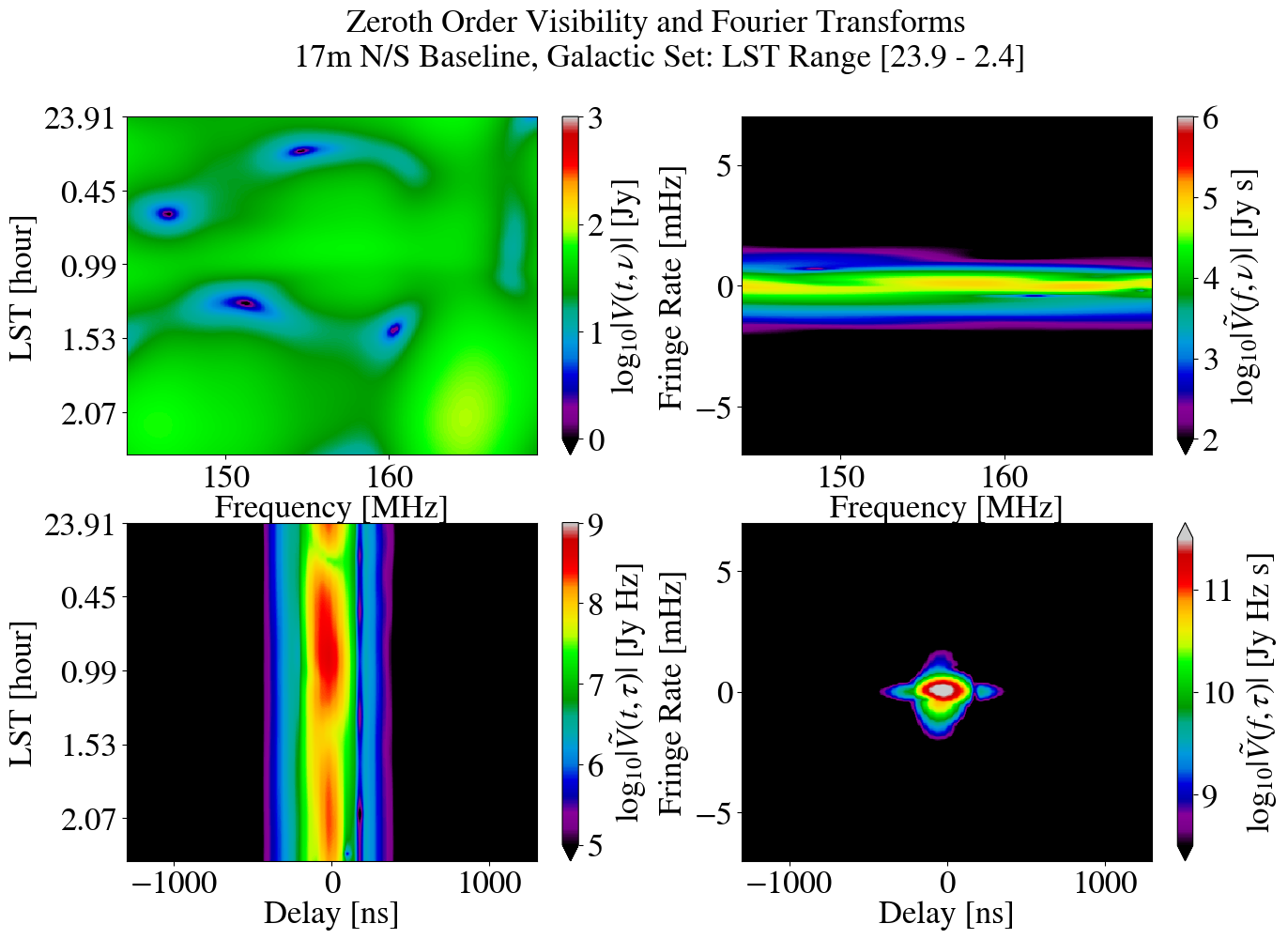}
    \caption{Zeroth-order visibility data of the combined (diffuse + point source) sky model, for a 17m N/S baseline, generated by the healvis simulator. Frequency range: 144-169 MHz. LST range: [2.4-4.9]. Moving clockwise from the top-left: time versus frequency ($t$ vs. $\nu$), fringe-rate versus frequency ($f$ vs. $\nu$), fringe-rate versus delay ($f$ vs. $\tau$), and time versus delay ($t$ vs. $\tau$).}
    \label{fig:zeroth_17m_NS_combined_GS}
\end{figure}

Figures (\ref{fig:zeroth_28m_EW_combined_GS}) and (\ref{fig:zeroth_17m_NS_combined_GS}) present the zeroth-order visibility of the combined sky model, for a 28m E/W baseline and a 17m N/S baseline, respectively. All possible Fourier transforms of the zeroth order visibilities are plotted. Moving clockwise from the top-left, the plots for each dataset are as follows: time versus frequency ($t$ vs. $\nu$), fringe-rate versus frequency ($f$ vs. $\nu$), fringe-rate versus delay ($f$ vs. $\tau$), and time versus delay ($t$ vs. $\tau$). The delay spectrum technique, originally explored in \citet{ParsonsBacker2009} and \citet{Parsons2012} is formed by Fourier transforming visibility data along the frequency axis, and is used extensively in the HERA collaboration to geometrically filter interferometric data; see, for example \citet{Kern2020a}. The fringe-rate analysis is formed by Fourier transforming visibility data along the time axis and is used extensively to filter interferometric data by temporal variation and sky region \cite{Parsons2016}. In Section (\ref{sec:PHENOMENON}), we will explain how first-order coupling features evolve as a function of apparent sky brightness and orientation, and how coupling manifests itself as a function of array position. To do so, we will use the fringe-rate versus delay (bottom right) Fourier transform as the main visual representation of coupling features.
\section{Simulation of First-Order Visibilities} \label{sec:FIRST_ORDER_SIM}
To aid in comparing the results of our simulation to data (in a future, companion paper), we choose to simulate the 143-element subset of the array (see Figure \ref{fig:Array_Plot}) where Vivaldi feeds and active analog system chains were fully commissioned. Of the hundreds of unique baseline groups which could be derived from this set of 143 elements, for brevity we limit our analysis to visibility data from 28m E/W baselines and 17m N/S baselines. These baselines are expected to contribute to HERA's primary scientific results and are of a convenient length to visually distinguish and differentiate various coupling features. The first-order simulations, which use as input the zeroth-order healvis visibility data, span the same frequency range of 144-169 MHz. Data from two different local sidereal time (LST) ranges will be studied. First, the LST range of [23.9-2.4], during which the galaxy sets in the western horizon with respect to HERA. Second, the LST range of [2.4-4.9], which captures the transit of Fornax A near the zenith point of HERA. In both LST ranges, coupling effects are measured down to at least 1 part in $10^{4}$ compared to peak visibility power. While our analysis focuses on how such coupling effects complicate the detection of the 21cm signal, the first-order coupling model we use to make such visibilities is noteworthy for any experiment using a radio interferometer where internal instrumental systematics at a level of 1 part in $10^{4}$ could corrupt scientific results.

\begin{figure*}
    \centering
    \includegraphics[width=1.0\textwidth]{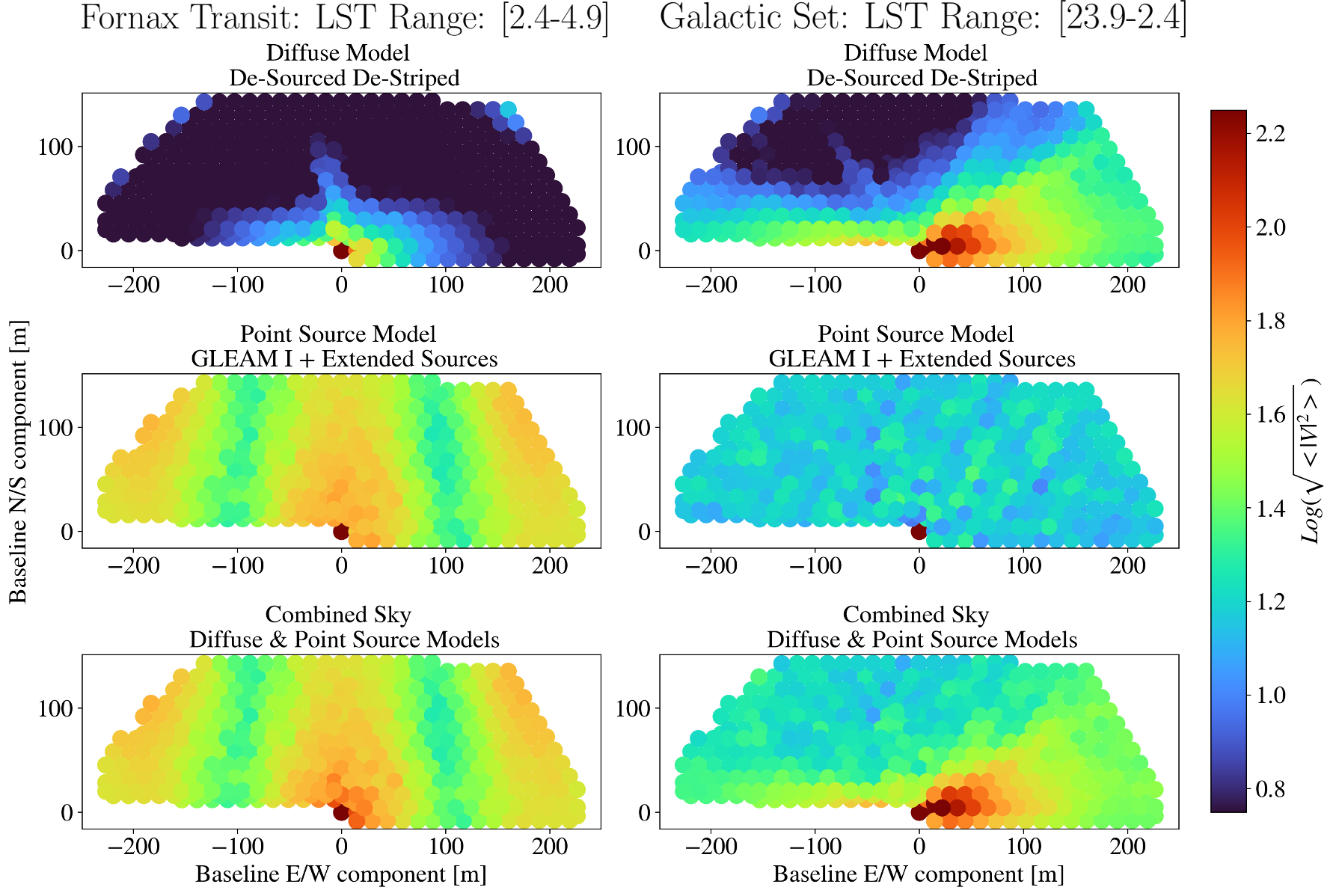}
    \caption{The RMS of the zeroth-order visibilities of all unique baseline groups in HERA, as a function of the E/W and N/S component of each respective group. Frequency range: [144-169] MHz. In the LST range [23.9-2.4], which corresponds to the galaxy setting in the horizon of HERA, baselines with short N/S components tend to be brighter in the diffuse sky model compared to the point source model, however the point source model dominates when the N/S component is longer. In the LST range [2.4-4.9], which captures Fornax A (a particularly bright point source) transiting the HERA beam, the HERA sky response at most (if not all) baselines is dominated by the point source model. By the method of stationary phase, discussed in Section (\ref{sec:PSPEC_wedge}), we note that the regions of the sky where the phase of an interferometric visibility is stationary (whether we are discussing the zeroth-order component, or the component associated with antenna-antenna scattering), occurs where the baseline vector aligns with the direction of strongest incident astrophysical radiation at the horizon. When the galaxy is setting, peak RMS power lies in mostly E/W baselines, which is the orientation of the galaxy in the horizon plane during LST range [23.9-2.4]. When Fornax A transits zenith, peak RMS power lies in mostly N/S baselines, which is the orientation of the galaxy in the horizon plane during LST range [2.4-4.9]}
    \label{fig:UVW_GS_FT}
\end{figure*}

\begin{figure*}
    \centering
    \includegraphics[width=0.81\textwidth]{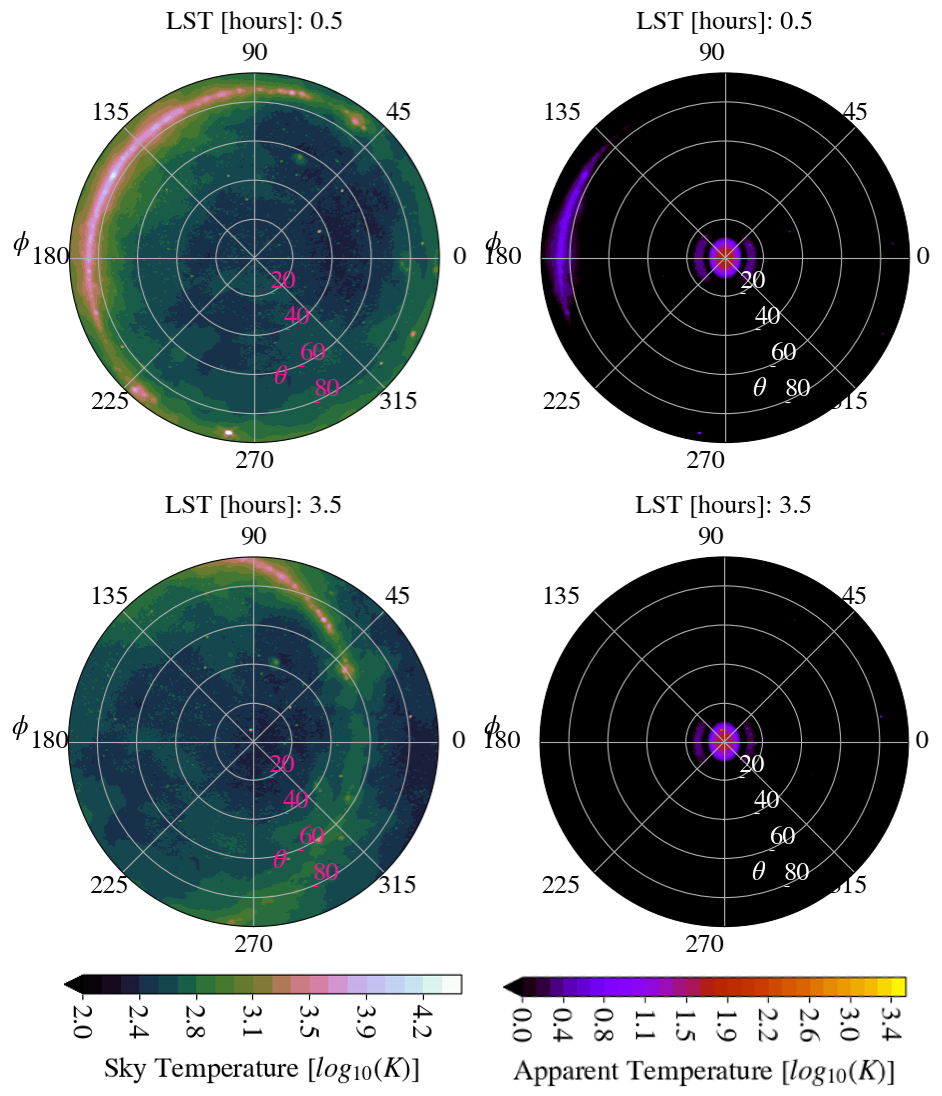}
    \caption{Simulations of the combined sky model model temperature and apparent temperature (sky multiplied by normalized beam gain) at 150 MHz, rotated into the coordinate system of the simulated HERA beam (see Figure \ref{fig:Spherical_Coordinates}) and plotted as an orthographic projection. At an LST of approximately 0.5, a significant fraction of the apparent brightness (namely the galactic plane) is concentrated near the horizon in the E/W direction. At an LST of approximately 3.5, Fornax A transits zenith, and a significant fraction of the apparent brightness is concentrated near zenith.}
    \label{fig:beam_apparent_GS_FT}
\end{figure*}

\section{Phenomenology}\label{sec:PHENOMENON}
\subsection{Using Fringe vs. Delay Space to Study Coupling Effects}
The structure of the general formalism suggests that fringe-rate versus delay is an ideal framework in which we can analyze first-order array element coupling effects. Recall from Section (\ref{sec:general_matrix_formalism}) that all terms associated with re-radiated (mutually-coupled) power introduce copies of zeroth-order visibilities. These copies are delayed by the time it takes for the reflected radiation to travel the coupled baseline vectors ($|\vec{b}_{ki}|$, $|\vec{b}_{kj}|$, depending on the sum term in Equation (\ref{eq:V_ij_1_with_0_simp})). Note also that, for each sum term in the formalism (either the simplified or general version), the only components which have dot products dependent on sky orientation (e.g. $\vec{b_{ij}}\cdot\hat{s}$) are the zeroth-order visibility terms. In (\ref{eq:V_ij_1_with_0_simp}), none of the other expressions are explicitly dependent on what sky is above the array, and so we infer that any time-dependent feature generated by first-order array element coupling must derive itself from the zeroth order visibility terms. 

As such, we interpret each component in the coupling sum terms of equation (\ref{eq:V_ij_1_with_0_simp}) as having one explicit fringe term. For each antenna $k$ included in $\sum_{k \neq i}$, we expect coupling features to manifest themselves at the dominant fringe rates of $\textbf{V}^0_{kj}$. Similarly, for $\sum_{k \neq j}$, we expect coupling features to manifest themselves at the dominant fringe rates of $\textbf{V}^0_{ik}$. The power of each coupling feature will be proportional to the RMS power of the respective visibilities at the simulated times and frequencies, e.g. $\sqrt{\langle{|\textbf{V}^0_{kj}|}^2\rangle}$ and $\sqrt{\langle{|\textbf{V}^0_{ik}|}^2\rangle}$. 

Figure (\ref{fig:UVW_GS_FT}) plots for two LST ranges the RMS power of the visibilities of all possible combinations of $(k,j)$ and $(i,k)$ in HERA, as a function of the E/W and N/S component of the baseline. Each colorful dot in the figures represents a unique baseline group in HERA, and the color of each dot represents the strength of the RMS visibility. The LST range [23.9-2.4] corresponds to the galaxy setting in the horizon of HERA. In this LST range, baselines with short N/S components tend to be brighter in the diffuse sky model compared to the point source model, however the point source model dominates when the N/S component is longer. The LST range [2.4-4.9] captures Fornax A, a particularly bright point source, transiting the main lobe of the HERA beam. In this LST range, the HERA sky response at most (if not all) baselines is dominated by the point source model.

\begin{figure}
    \centering
    \includegraphics[width=.5\textwidth]{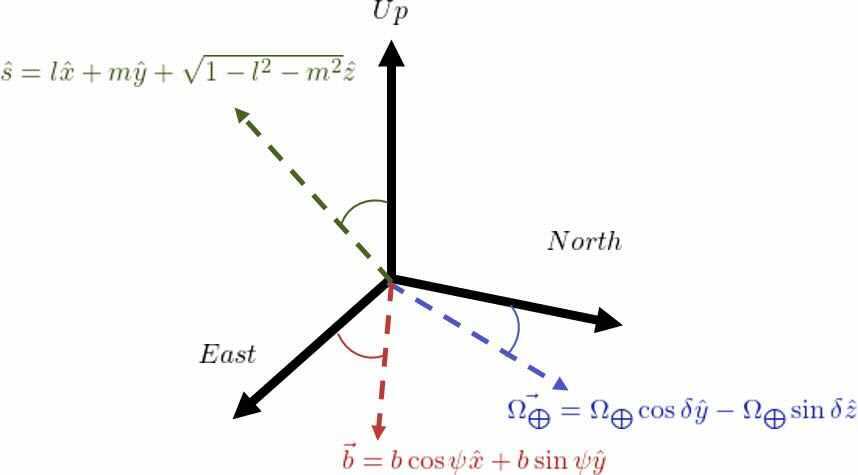}
    \caption{Geometry used to compute coordinate expression for fringe-rate as a function of sky-location and baseline length and location in equation~\ref{eq:FRINGERATE}}
    \label{fig:FR_COORDS}
\end{figure}

\begin{figure*}
    \centering
    \includegraphics[width=1.0\textwidth]{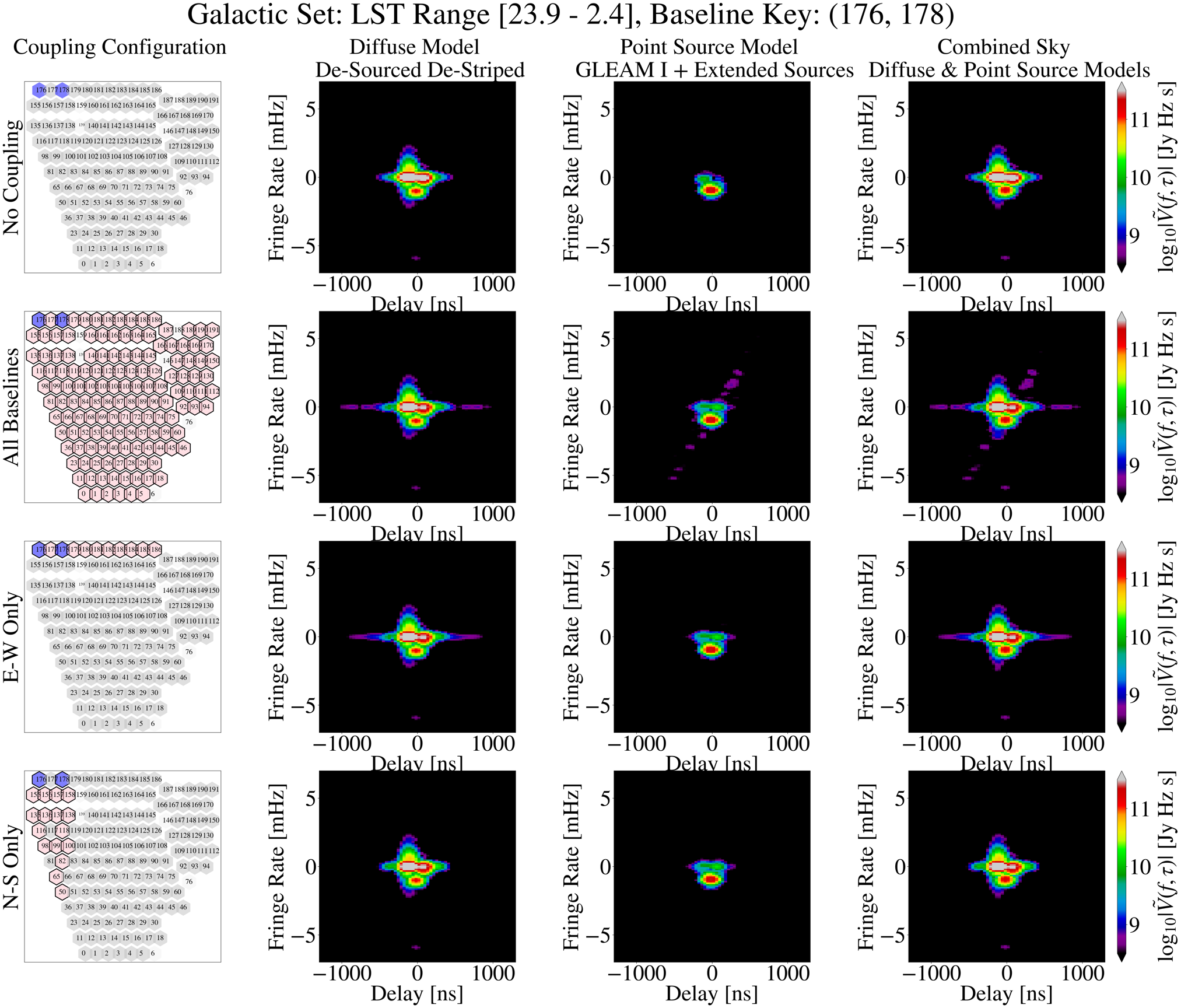}
    \caption{Fringe rate versus delay plot of the first-order visibility of 28m E/W baseline (176, 178) of the HERA array, plotted in blue, as well all other elements (grey) which were installed on JD 2459122. Three different array configurations are plotted: 'All Baselines', 'E-W Only', and 'N-S Only'. For each array configuration, only antennae which are plotted in pink are included in the first-order coupling calculation (equation \ref{eq:V_ij_1_with_0_simp}). All elements of the simulated array configuration use identical beam products ($\textbf{J}(\hat{s})$, $\textbf{J}^{\dagger}(\hat{s})$). LST range: [23.9-2.4], Frequency range: [144-169 MHz]. In this LST range, the galaxy sets in the western horizon with respect to HERA. See Figure (\ref{fig:beam_apparent_GS_FT}) for sky brightness as a function of position in the HERA beam.}
    \label{fig:176_178_GS}
\end{figure*}

\begin{figure*}
    \centering
    \includegraphics[width=1.0\textwidth]{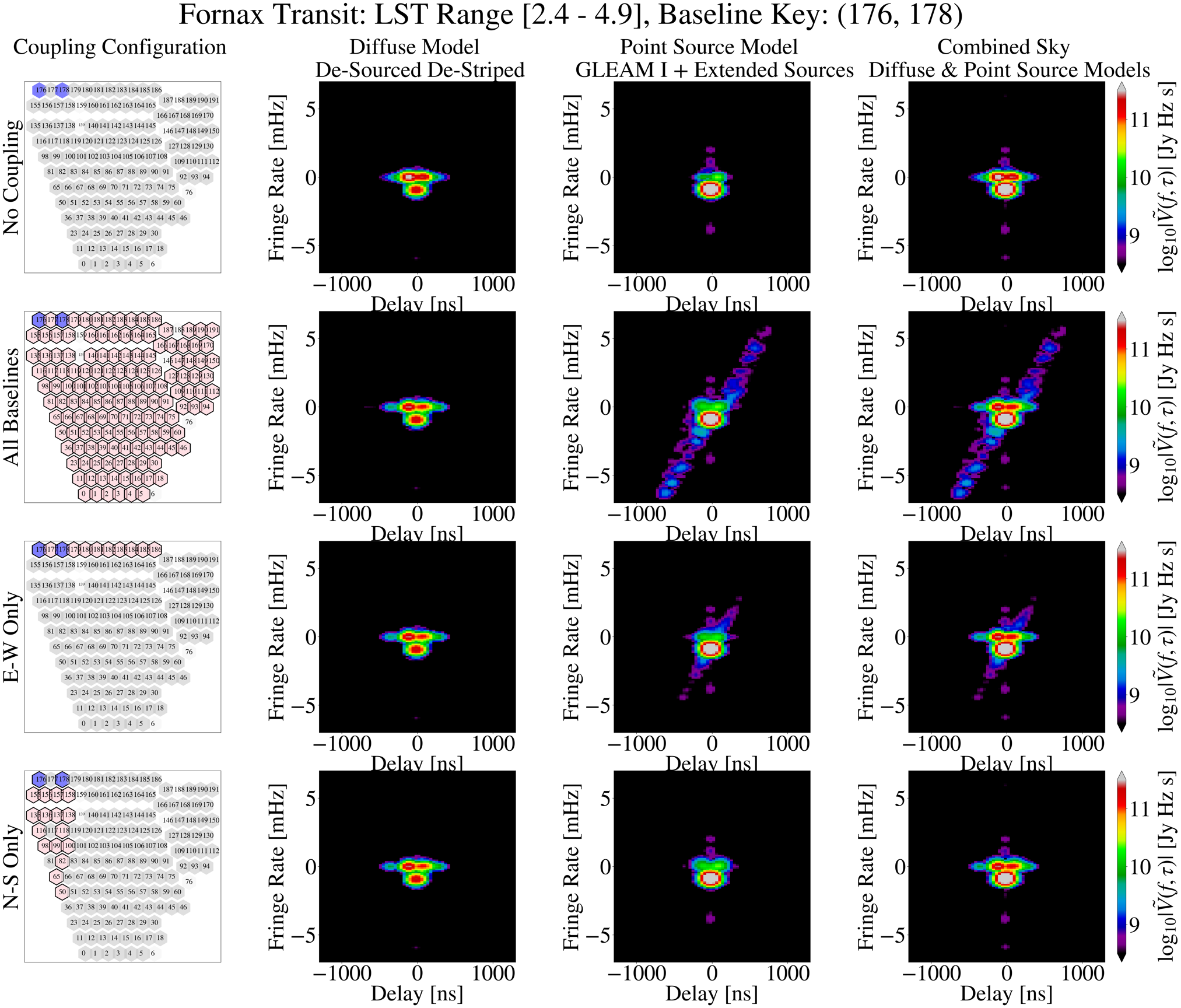}
    \caption{Fringe rate versus delay plot of the first-order visibility of 28m E/W baseline (176, 178) of the HERA array, plotted in blue, as well all other elements (grey) which were installed on JD 2459122. Three different array configurations are plotted: 'All Baselines', 'E-W Only', and 'N-S Only'. For each array configuration, only antennae which are plotted in pink are included in the first-order coupling calculation (equation \ref{eq:V_ij_1_with_0_simp}). All elements of the simulated array configuration use identical beam products ($\textbf{J}(\hat{s})$, $\textbf{J}^{\dagger}(\hat{s})$). LST range: [2.4-4.9], Frequency range: [144-169 MHz]. In this LST range, Fornax A transits near the zenith point (location of peak gain) of the HERA beam. See Figure (\ref{fig:beam_apparent_GS_FT}) for sky brightness as a function of position in the HERA beam.}
    \label{fig:176_178_FT}
\end{figure*}

\subsection{Delay Contributions of First-Order Coupling} \label{sec:phenomenon_delay_contribs}
 Each sum term in equation (\ref{eq:V_ij_1_with_0_simp}) contains four complex expressions which contribute to the delay at which first-order coupling will manifest itself: First, the constant values (e.g. $i\eta_{0}\Gamma$). Second, the beam product (e.g. $J(\hat{b_{ik}})J(\hat{b_{ki}})$). Third, delays associated with the copied visibilities (e.g. $V^0_{kj}$). Finally, the complex exponential associated with baseline length (e.g. $e^{ 2\pi i \frac{v}{c}|\vec{b_{ki}}|}$). Regarding the constants, we have fit our voltage reflection coefficient measurement using high-order Chebyshev polynomials, thus preserving its spectral shape while ensuring any artifacts in delay space are negligible. We can confirm that all constant values contribute negligible power at nonzero delays. 

\begin{figure}
    \centering
    \includegraphics[width=.47\textwidth]{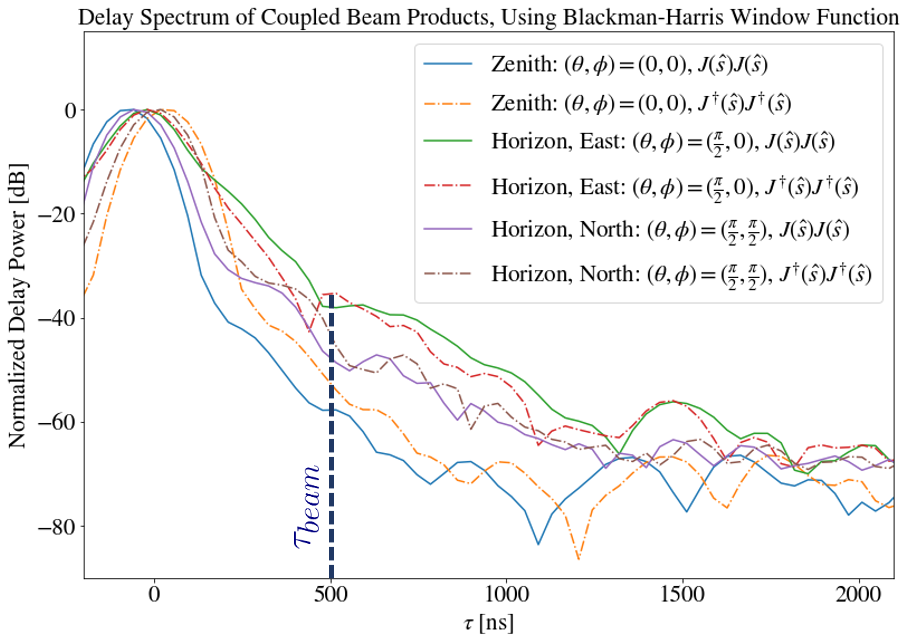}
    \caption{A peak-normalized delay spectrum of different healpix beam pixels, caused by the window function applied to the beam products of our first-order visibilities, $J({\hat{s}})J(\hat{s})$ (solid lines) and $J^{\dagger}(\hat{s})J^{\dagger}(\hat{s})$ (dashed lines). Window function used: Blackman-Harris. The plotted pixels chosen correspond to zenith (blue/orange) and two orthogonal positions in the horizon plane (green/red and purple/burgundy, respectively). For both beam products, regardless of pixel location, there is power associated with the beam at non-negligible delays. For all pixel positions, power is below -40dB of the peak-normalized window function power after $|\tau| \geq 500ns$. This nonzero delay cannot be ignored when analyzing first-order visibilities.  One may see this beam delay feature manifest itself in first-order power spectra. See, for example, $\tau_{beam}$ in Figure (\ref{fig:PSPEC_WEDGE_DESCRIBED}). }
    \label{fig:delay_of_beam_window_func}
\end{figure}

To gain intuition on the third contribution, Figure (\ref{fig:delay_of_beam_window_func}) plots the peak-normalized delay spectrum of different healpix beam pixels, caused by the window function applied to the beam products of our first-order visibilities, $J({\hat{s}})J(\hat{s})$ (solid lines) and $J^{\dagger}(\hat{s})J^{\dagger}(\hat{s})$ (dashed lines). Whenever performing a Fourier Transform of a beam product, whether in healvis, or in the first-order coupling algorithms, we use a Blackman-Harris window. \citet{Nithya2016_Blackman_Harris} studied the leakage of foreground power into the time domain associated with windowing in the frequency domain and concluded that the Blackman-Harris window offers a contamination-free dynamic range of at least 1 part in $10^6$ for voltage signals in the time domain. Thus, foreground leakage caused by the window will be below the expected EoR signal, although to the detriment of measurement sensitivity, which is reduced by approximately 50 percent. The healpix beam pixels plotted in Figure (\ref{fig:delay_of_beam_window_func}) are at zenith (blue/orange) and two orthogonal horizon positions (green/red and purple/burgundy, respectively). Power is associated with the windowed beam function at non-negligible delays in both beam products, regardless of pixel location. Nevertheless, for all pixel positions, this window function power is below -40dB of the peak-normalized delay spectra after $|\tau| \geq 500ns$. Since power drops off so quickly at delays greater than $500ns$, regardless of pixel, we approximate the power associated with beam products to be approximately equal across all pixels. Similarly, it can be shown that the Blackman-Harris window function applied to the zeroth-order beam product, $J(\hat{s})J^{\dagger}(\hat{s})$, produces a comparable delay spectrum to those plotted in Figure (\ref{fig:delay_of_beam_window_func}). Thus, we approximate the "delay spectrum associated with the window function of the beam" as equal across all pixels and all beam products:
\begin{equation}\label{eq:tau_beam}
\tau_{J(\hat{s})J^{\dagger}(\hat{s})} \approx  \tau_{J(\hat{s})J^(\hat{s})} \approx \tau_{J^{\dagger}(\hat{s})J^{\dagger}(\hat{s})} \equiv \tau_{beam} \approx 500ns
\end{equation}
$\tau_{beam}$ is typically ignored when discussing the delays associated with power in zeroth-order visibilities. However, this nonzero beam delay term cannot be ignored when analyzing first-order visibilities.  Section (\ref{sec:PSPEC_wedge}) describes how terms of $\tau_{beam}$ manifest themselves in the structure of first-order coupling effects. These beam delay terms are conceptually labelled atop the 2D power spectrum measurement presented in Figure (\ref{fig:PSPEC_WEDGE_DESCRIBED}).

Regarding the fourth delay contribution, we note that zeroth-order visibilities $V^0_{kj}$ tend to have maximum power at two different delays: those associated with the primary beam lobe and those associated with the 'geometric horizon' of the baseline ($|\vec{b_{kj}}|$). The manifestation of foregrounds (the primary source of incident astrophysical radiation) in delay space, as a function of interferometric baseline length, has been studied extensively. See, for example, \citet{WEDGE_DATTA_2010}, \citet{Parsons2012}, \citet{WEDGE_MORALES_2012}, and \citet{WEDGE_NITHYA_2015}. Such studies describe how foregrounds of wideband antenna arrays are contained in a 'wedge' of delay versus baseline length (see, for example, the $V^0_{ij}$ subplots of Figures (\ref{fig:PSPEC_WEDGE_GS}) and (\ref{fig:PSPEC_WEDGE_FT}), and the description of the 'wedge' in Figure (\ref{fig:PSPEC_WEDGE_DESCRIBED}). In this wedge, power appears at delays approximately equal to zero, regardless of baseline length. Such power comes from incident astrophysical radiation being absorbed in the direction of maximum beam gain. The wedge also contains power at delays which are directly proportional to the baseline length of the measured visibility, with a slope which is inversely proportional to the speed of light. \citet{WEDGE_NITHYA_2015} describes this "pitchfork" effect at the geometric horizon of the array as fundamental to interferometric visibilities measured by wide-band array elements;  this is not an instrumental systematic effect, but a fundamental instrument response. As such, the effect can only be mitigated, not entirely removed. \citet{Kern2020a} describes how various instrumental effects differ in fringe and delay space compared to the pitchfork effect.

Knowing that zeroth-order visibilities $(V^0_{kj})$ have significant power at two separate delays ($\tau \approx 0, \tau \approx \frac{|\vec{b_{kj}}|}{c}$), and recalling the fourth contribution to total delay (complex exponentials of the form $e^{2\pi i \frac{v}{c}|\vec{b_{ki}}|}$), we conclude that first-order coupling effects of the ${ij}^{th}$ visibility predominately reside at the following delays:

\begin{equation}\label{eq:tau_coupling}
    \tau^1_{ij} \approx 
    \begin{cases}
        \tau_{beam} \rightarrow V^0_{ij} \hspace{0.1in} "wedge"\\
        \tau_{|\vec{b_{ij}}| } \rightarrow  V^0_{ij} \hspace{0.1in} "wedge"\\
        \tau_{|\vec{b_{ki}}| } + \tau_{   V^0_{kj, main\hspace{0.05cm} lobe} } + \tau_{beam}  \\
        \tau_{|\vec{b_{ki}}| } + \tau_{   V^0_{kj, geometric\hspace{0.05cm} horizon} } + \tau_{beam}
    \end{cases}
\end{equation}

While zeroth-order visibilities also contain significant power at delays associated with $\tau_{beam} $ and $\tau_{|\vec{b_{ij}}| }$, the latter two cases of equation (\ref{eq:tau_coupling}) are unique to first-order coupling. The maximum extent in delay of these latter two cases is plotted in yellow in Figure (\ref{fig:PSPEC_WEDGE_DESCRIBED}). 
\begin{figure}
    \centering
    \includegraphics[width=.37\textwidth]{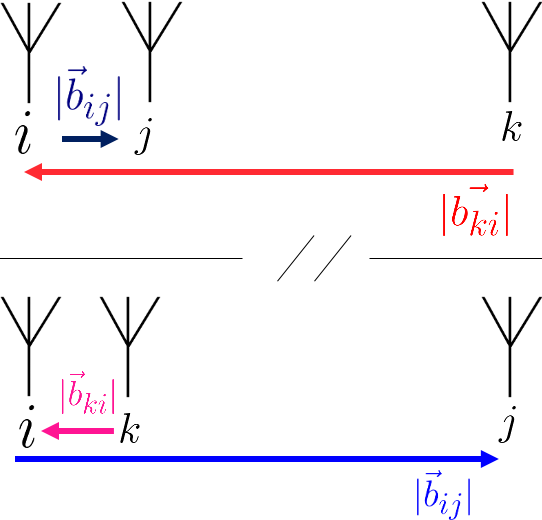}
    \caption{Two instances of antenna-antenna coupling terms, which help describe the inverse wedge phenomenon outlined in Figure (\ref{fig:PSPEC_WEDGE_DESCRIBED}). Suppose we are analyzing the first-order coupling of the $ij^{th}$ baseline, $V^1_{ij}$, as formalized in equation (\ref{eq:V_ij_1_with_0_simp}). This visibility will contain copies of all other visibilities, at a delay proportional to the respective baseline distances, $|\vec{b_{ki}}|$ or $|\vec{b_{jk}}|$. When the baseline of the measured visibility, $|\vec{b_{ij}}|$, is shortest, we maximize the possible length of $|\vec{b_{ki}}|$ and $|\vec{b_{jk}}|$. Similarly, when the baseline of the measured visibility, $|\vec{b_{ij}}|$, is longest, we minimize the possible length of $|\vec{b_{ki}}|$.}
    \label{fig:ki_decrease_ij_increase}
\end{figure}

\subsection{Fringe-Rate as a Function of Orthographic Sky Location}\label{app:fr_derivation}
In this section, we use the coordinate-free fringe-rate expression from \citet{Parsons2016} to derive a useful analytic formula relating fringe rate to orthographic-projected sky location, baseline length, and baseline orientation. Doing so offers us intuition regarding the fringe-rates associated with zeroth-order visibility terms ($\scriptscriptstyle{\textbf{V}^0_{kj}}$ and $\scriptscriptstyle{\textbf{V}^0_{ij}}$). 

According to \citet{Parsons2016},
\begin{equation}
    f_r \approx \frac{\nu}{c} \left[ \boldsymbol{\hat{s}} \cdot (\boldsymbol{\Omega_{\oplus}} \times \boldsymbol{b}) \right]
\end{equation}

This geometry can be translated into the ground-based orthographic coordinate system illustrated in Figure (\ref{fig:FR_COORDS}) by aligning the $\boldsymbol{\hat{x}}$ axis with the E/W direction, the $\boldsymbol{\hat{y}}$ axis with the the N/S direction, and the $\boldsymbol{\hat{z}}$ axis with the vertical (zenith) direction. If our array is located at a declination of $\delta$ and our baseline is at an angle of $\psi$ with respect to the E/W direction, then 
\begin{equation}
    \boldsymbol{\Omega_\oplus} \times \boldsymbol{b} =  b \left[ \boldsymbol{\hat{x}} \sin \delta \sin \psi - \boldsymbol{\hat{y}} \sin \delta \cos \psi + \boldsymbol{\hat{z}} \cos \psi \cos \delta \right]
\end{equation}
Writing $\boldsymbol{\hat{s}}$ in orthographic coordinates, 
\begin{equation}
    \boldsymbol{{\hat{s}}} = \boldsymbol{\hat{x}} \ell + \boldsymbol{\hat{y}} m + \boldsymbol{\hat{z}}  \sqrt{1 - \ell^2 - m^2} 
\end{equation}
we have
\begin{align}\label{eq:FRINGERATE}
    f_r(\ell, m) = \frac{b \nu}{c} \bigg( \ell \sin \delta \sin \psi - m & \sin \delta \cos \psi \\
    & + \sqrt{1 - \ell^2 - m^2} \cos \psi \cos \delta \bigg) \nonumber
\end{align}
When the majority of apparent brightness is concentrated at zenith, power is concentrated at fringe rates of $f_r(0, 0) \approx \frac{b\nu}{c} \cos \psi \cos \delta$, which is proportional to the baseline's projected E/W length ($b \cos \psi$). The same is true when apparent brightness is concentrated near the horizon in the N/S direction. On the other hand, when apparent brightness is concentrated near the horizon, in the E/W direction, then the peak fringe rate is proportional to the N/S projected length of the baseline $f_r(1, 0) \propto \frac{b\nu}{c} \sin \psi \cos \delta$.

\subsection{Fringe-Rate Filtering: A Potential Strategy for Mitigating First-Order Coupling Effects} \label{sec:frf}
Recall from equation (\ref{eq:V_ij_1_with_0_simp}) that first-order visibilities consist of a sum of copies of zeroth order visibilities, and that these visibilities ($V^0_{kj}$ or $V^0_{ik}$, depending on the sum term) are the only fringing component in each respective sum. In the first sum, visibilities will have fringe rates associated with $\vec{b_{kj}}, k \neq i$. The fringe terms associated with the main beam of the coupled visibility ($\frac{|\vec{b_{kj}|}}{c}\cos \psi \cos \delta$, or  $\frac{|\vec{b_{kj}|}}{c}\sin \psi \cos \delta$) will equal the fringe term associated with apparent brightness being near zenith of the measured baseline ($\frac{|\vec{b_{ij}|}}{c}\cos \psi \cos \delta$) when the E/W component of $|\vec{b_{kj}}|$ equals the E/W component of $|\vec{b_{ij}}|$. Similarly, for the second sum, the fringe terms associated with the main beam of the coupled visibility ($\frac{|\vec{b_{ik}|}}{c}\cos \psi \cos \delta$, or $\frac{|\vec{b_{ik}|}}{c}\sin \psi \cos \delta$) will equal the fringe term associated with apparent brightness being near zenith of the measured baseline ($\frac{|\vec{b_{ij}|}}{c}\cos \psi \cos \delta$) when the E/W component of $|\vec{b_{ik}}|$ equals the E/W component of $|\vec{b_{ij}}|$. In our simulated HERA-like array, these relationships are infrequently satisfied. Thus, the fringe rates associated with first-order antenna-antenna coupling will not typically be the same fringe rates associated with the main beam of the measured visibility. Evidence for this can be found in all fringe-rate versus delay plots presented in this work (Figures (\ref{fig:176_178_GS}), (\ref{fig:176_178_FT}), and all plots in the online supplementary material). To the dynamic range of each figure, the non-fringing and fringing features associated with first-order coupling appear at fringe-rates different than the main lobe for all baselines with a nonzero E/W component. This is true for all analyzed baseline lengths at all simulated LST ranges. While it is beyond the scope of this work to test filtering strategies for first-order coupling systematics, our conclusion regarding main lobe fringe rates may prove crucial for finding subsets of fringe versus delay space which are not contaminated by mutual coupling.

Let us now broadly categorize coupling features as being non-fringing or having non-zero fringe, and study the two cases separately.

\subsection{Non-Fringing ("Bar") Features}
The top row of Figure (\ref{fig:beam_apparent_GS_FT}) shows an LST in our combined sky simulation where a significant portion of the apparent sky brightness is concentrated near the horizon, in the E/W direction. In these figures, the sky has been rotated into an orthographic projection of the spherical coordinate system used for our beam, as presented in Figure (\ref{fig:Spherical_Coordinates}). Thus, all plots in Figure (\ref{fig:beam_apparent_GS_FT}) depict the sky as it is 'seen' by a HERA beam element as it receives incident astrophysical radiation. At an LST of 0.5 (top row), the galaxy is setting in the western horizon with respect to HERA. Figure (\ref{fig:176_178_GS}) records the first-order coupling ($\textbf{V}^1_{ij}$) of baseline (176, 178) - plotted in blue - as calculated by (\ref{eq:V_ij_1_with_0_simp}), in a range of LSTs including 0.5. This coupling term is calculated for all three different sky models (diffuse only, point source only, and the combined sky) for three unique array configurations. For each array configuration, only antennae which are plotted in pink in column 1 are included in the first-order coupling calculation. In addition to the three unique array configurations (rows 2-4), the first row of Figure (\ref{fig:176_178_GS}) plots the zeroth-order visibility of the same sky models (row 1).

The brightness of the galactic plane is a feature which is only present in the diffuse sky model, not the point source model. Therefore, for all array configurations in column 2 of Figure (\ref{fig:176_178_GS}), the apparent brightness is concentrated near the horizon in the E/W direction. As predicted by (\ref{eq:FRINGERATE}), the peak fringe rate of coupling features is proportional to the N/S projected length of the coupled (pink) baselines. For the 'E/W Only' array configuration (row 3), all coupled baselines have a N/S component of zero with respect to the antennae in baseline (176, 178). Thus, coupling manifests itself at a fringe rate of zero over a broad range of delays, creating a bar-like feature. This bar-like feature is more prominent in the 'All Baselines' configuration (row 2, column 2) because there are more coupled antennae which have a non-zero E/W baseline component. If a main-beam fringe-rate filtering strategy were adopted, such as that suggested in Section (\ref{sec:frf}), our phenomenological understanding of bar-like features suggests we ignore N/S baselines entirely from our analysis if the array is two-dimensional (i.e. not all baselines lie in the N/S direction).

By contrast, all coupled antennae in the 'N/S only' array (row 4) of column 2 have a negligible E/W component with respect to (176, 178), and so instead of producing a bar feature, the coupled antennae will manifest themselves in baseline (176, 178) with nonzero fringe. At the dynamic range of the colorbar of this figure, this is not apparent, both because the gain in the N/S direction for this simulated beam polarization (the N/S oriented dipole of the Vivaldi feed) is significantly lower than the gain of that feed in the E/W direction. Further, Figure (\ref{fig:UVW_GS_FT}) tells us that these fringing N/S baseline contributions will have less RMS power in their visibilities than the RMS of the visibilities which contribute to the 'E/W' array configuration. Thus, for this LST range and array configuration, features with non-zero fringe are subordinate to zero-fringe bar features. For the rest of this work, subordinate coupling features are not presented or discussed. 

In contrast to the diffuse sky model, the point source sky model (column 3) in Figure (\ref{fig:176_178_GS}) depicts coupling features that dominate at non-zero fringe rates. We note from Figure (\ref{fig:UVW_GS_FT}) that the RMS power of the point source sky at this LST range is fairly homogeneous regardless of baseline length or orientation. Thus, we may treat the apparent brightness as coming from the pointing center of the HERA beam, which is zenith. The 'All Baselines' array configuration of Figure \ref{fig:176_178_GS} (row 2, column 3) has a sufficient number of coupled antennae with a nonzero E/W component that we begin to see structure with non-zero fringe rates. Coupling features with non-zero fringe rates are further discussed in the next section.

\subsection{Fringing ("Cross") Features}
\label{sec:cross_features}
The bottom row of Figure (\ref{fig:beam_apparent_GS_FT}) shows an LST in our combined sky simulation where a significant fraction of the apparent brightness is concentrated near zenith. From Figure (\ref{fig:UVW_GS_FT}), we know that the HERA response of most (if not all) baselines in this LST range is dominated by the point source model, not the diffuse model. Fornax A transits the zenith point of the HERA array at an LST of approximately 3.5. Figure (\ref{fig:176_178_FT}) records the first-order coupling ($\textbf{V}^0_{ij}$), of baseline (176, 178) as calculated by (\ref{eq:V_ij_1_with_0_simp}), over a range of LSTs, including 3.5. The brightness of Fornax A is a feature which is only present in the point source sky model, not the diffuse sky model. Thus, for all array configurations in column 3 of Figure (\ref{fig:176_178_FT}), there is a significantly bright source at zenith. Therefore, per equation (\ref{eq:FRINGERATE}), the peak fringe rate for first-order coupling features will be proportional to the E/W projected length of the coupled (pink) baselines. For the "E/W Only" array configuration (row 3), all coupled antennae have a non-zero E/W component with respect to the antennae in baseline (176, 178) and coupling manifests itself at non-zero fringe rates over a broad range of delays, creating a cross-like feature. The cross feature is weaker in the "E/W Only" configuration compared  to the "All Baselines" configuration because the latter contains more E/W baselines. We note from the "N/S Only" configuration that purely N/S baselines do not contribute to the cross feature when the apparent brightness is concentrated at zenith.

The slope of this cross-like coupling feature in fringe vs. delay space depends on the relative position of the baseline with respect to all other antennae in the array. In baseline (176, 178), coupled antennae are physically located to the east of the baseline, which, due to the direction of the complex exponential terms in (\ref{eq:V_ij_1_with_0_simp}) makes coupled antennae manifests themselves in the first order visibility of baseline (176, 178) with positive fringe space at positive delay, and with negative fringe rates at negative delay (see row 2, column 3 of Figure (\ref{fig:176_178_FT})). In the same simulation and array configuration of baseline (128, 130) - row 2, column 3 of Figure (S6) of the online supplemental material - coupled antennae are physically located to the west of the baseline, making the coupling manifest itself with opposite slope compared to baseline (176, 178). By extension, the coupling features of an identical baseline in the middle of the array, for example (162, 164), having coupled antennae on both sides of the observed baseline, will manifest itself along both slopes, making a characteristic 'X' feature (see Figure (S2) of the online supplemental material). It is worth clarifying that the 'X' feature is purely a consequence of the geometric location of the baseline with respect to the other antennae in the array, not the fact that the baseline happens to be oriented in the E/W direction. Figure (S4) of the online supplemental material depicts a 17m N/S baseline in the middle of the array, (143,163), where first-order coupling also manifests itself as an 'X'.  

To the dynamic range of all fringe vs. delay plots presented in this work, we note that cross-like coupling systematics do not appear at fringe-rates that correspond to the main beam of the measured visibility. This is further evidence that, in a HERA-like redundant array, the E/W component of $|\vec{b_{kj}}|$ and $|\vec{b_{ik}}|$ rarely equals the E/W component of $|\vec{b_{ij}}|$. Thus, a less-contaminated analysis space, per baseline group, may reside at fringe rates associated with the main beam of baselines which have a nonzero E/W component. This could be critical to any future filtering strategy used to mitigate antenna-antenna coupling systematics.

\subsection{Mutual Coupling: A Threat to Redundant Baseline Averaging}
Three unique 28m E/W baselines are plotted in the following combinations of two figures: Figures (\ref{fig:176_178_GS} and \ref{fig:176_178_FT}) of this work, and Figures (S1 and S2), and (S5 and S6) of the online supplemental material. The zeroth-order (row 1, 'No Coupling') visibilities of all sky models (columns 2-4) of each of these 28m baselines are identical. As such, without coupling these baselines would be considered 'redundant', meaning they could be averaged together in the complex plane without distorting information about incident astrophysical radiation, because their visibilities are identical for all frequencies and times (or, analogously, all delays and fringe rates). However, when we include first-order coupling, these visibilities are no longer identical. First-order coupling makes baselines which were otherwise redundant no longer have identical sky responses. For redundantly-configured arrays such as HERA to achieve the measurement sensitivity required to make interferometric measurements of the EoR (the principal scientific goal of the experiment) in any practical duration of observation time, it is imperative, according to analyses such as those presented in  \cite{2020MNRAS.499.5840D}, that visibilities from redundant baselines be able to be redundantly averaged - for at least $\emph{some}$ subset of fringe and delay space. It is beyond the scope of this paper to further address the challenge of non-redundancy caused by array element coupling, but a follow-up study could investigate filtering strategies whereby first-order visibilities of redundant baselines may continue to be identical in certain subsets of fringe and delay space, at least to the level of EoR sensitivity.

\section{Effects of First-Order Coupling on Power Spectrum Measurements}
\label{sec:PSPEC}
First-order coupling has a significant effect on measurements of the power spectrum, which complicate the fundamental foreground mitigation strategies currently used in 21cm cosmological analyses. We calculate power spectra of our first order visibility data using the formalism of \citet{Parsons2012}, as normalized using the quadratic estimator strategy of \citet{2020heraPP}, and instantiated into code using the hera$\_$pspec power spectrum estimator \footnote{https://github.com/HERA-Team/hera$\_$pspec}. To construct the power spectra, we combine information which is redundant both in time and baseline ENU coordinates. Pairs of baselines within each respective redundant baseline group are constructed to estimate a power spectrum. For example, the redundant group of 28m E/W baselines would contain the following pairs: ((176, 178, "xx"), (176, 178, "xx")) and ((162, 164, "xx"), (176, 178, "xx")). For reference, these baselines are plotted for two different LST ranges in Figures (\ref{fig:176_178_GS}), (\ref{fig:176_178_FT}), and Figures (S1) and (S2) of the online supplemental material. Power spectrum measurements which are constructed using "auto" baseline pair combinations contain only those combinations of baseline pairs within a redundant group of one pair with itself [e.g. ((176, 178, "xx"), (176, 178, "xx"))]. Data sets which are "cross"-constructed contain all other possible combinations of baseline pairs in the same redundant group, excluding auto baseline pairs [e.g. ((162, 164, "xx"), (176, 178, "xx")), but not ((162, 164, "xx"), (162, 164, "xx"))]. 

Regardless of whether the power spectrum is estimated using auto baseline pair or cross baseline pair combinations, we may at our convenience average the data across time (i.e. one power spectrum per time) and/or redundant baseline groups (i.e. one power spectrum per baseline group). We will analyze data which has been averaged along one, and then both, of these axes. First, we study the effects of first-order coupling on auto-correlation data [e.g. $V_{ii}$ or $V_{jj}$], regardless of if the power spectrum measurement was constructed using auto baseline pair or cross baseline pair combinations. Next, we will average across both time and redundant baseline group, and analyze in one figure the power spectra of all possible baselines in the array, as a function of delay.

\subsection{Auto-Correlation Data: A "Real" Coupling Effect}\label{sec:PSPEC_autocorr}

\begin{figure*}
    \centering
    \includegraphics[width=.81\textwidth, height=15.1cm]{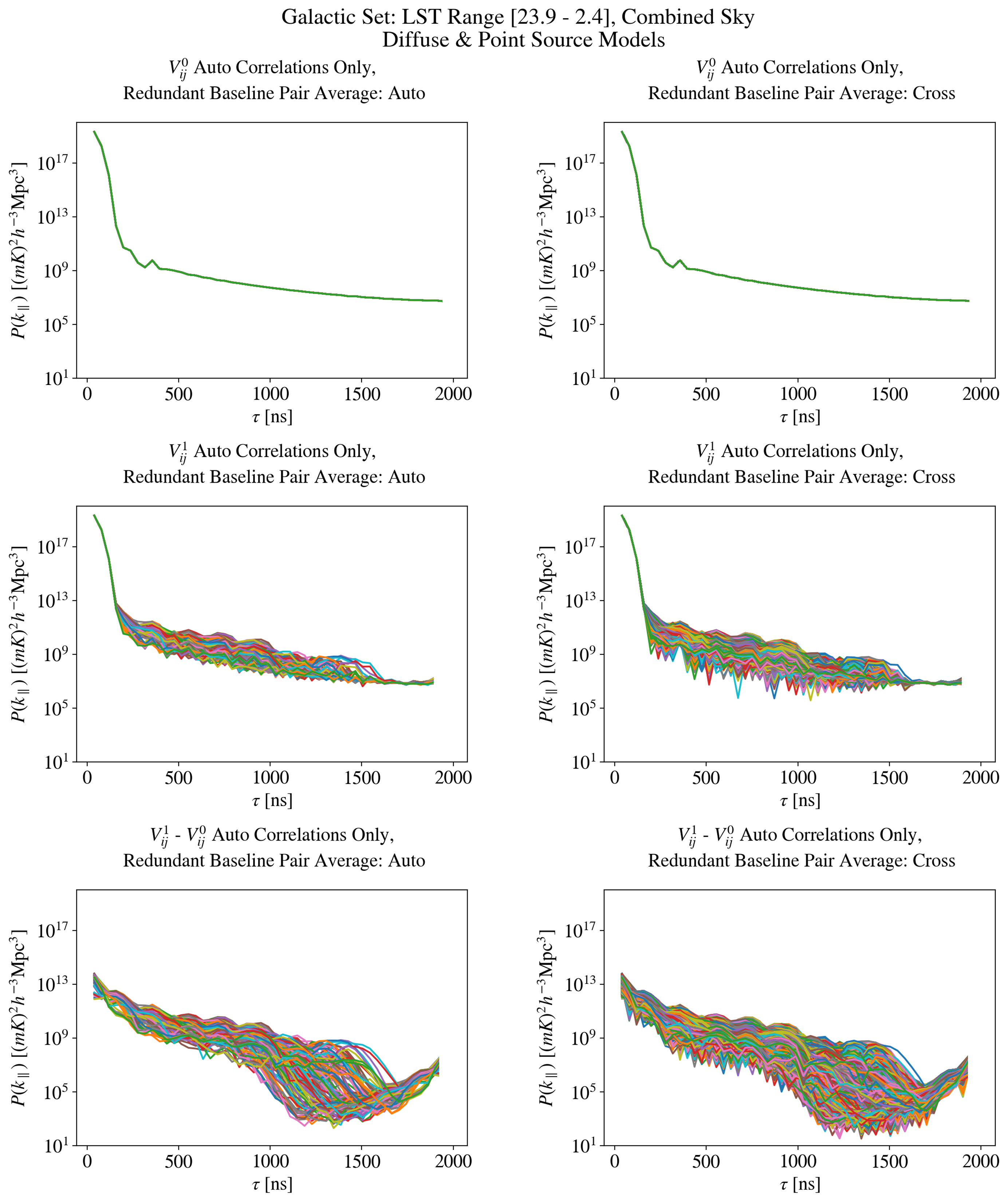}
    \caption{Time-averaged (but not baseline-averaged) power spectrum measurements of auto-correlation data, as a function of delay. The data products used in this plot are a pair of baselines, each belonging to the same redundant group. For example, this redundant group contain auto-correlations, e.g. ((176, 176, 'xx'), (176, 176, 'xx')) or ((162, 162, 'xx'), (178, 178, 'xx')). Data sets with an 'auto' baseline pair construction only contain baselines paired with itself [e.g. ((176, 176, 'xx'), (176, 176, 'xx'))]. Data sets with a 'cross' baseline pair construction contain baseline pairs with all other baseline pairs in the same redundant group, except for itself [e.g. ((162, 162, 'xx'), (178, 178, 'xx'))]. Frequency range: [144-169] MHz. LST range: [23.9-2.4], which corresponds to the galaxy setting in the horizon of HERA.}
    \label{fig:PSPEC_AUTO_BLP_GS}
\end{figure*}

\begin{figure*}
    \centering
    \includegraphics[width=.81\textwidth, height=15.1cm]{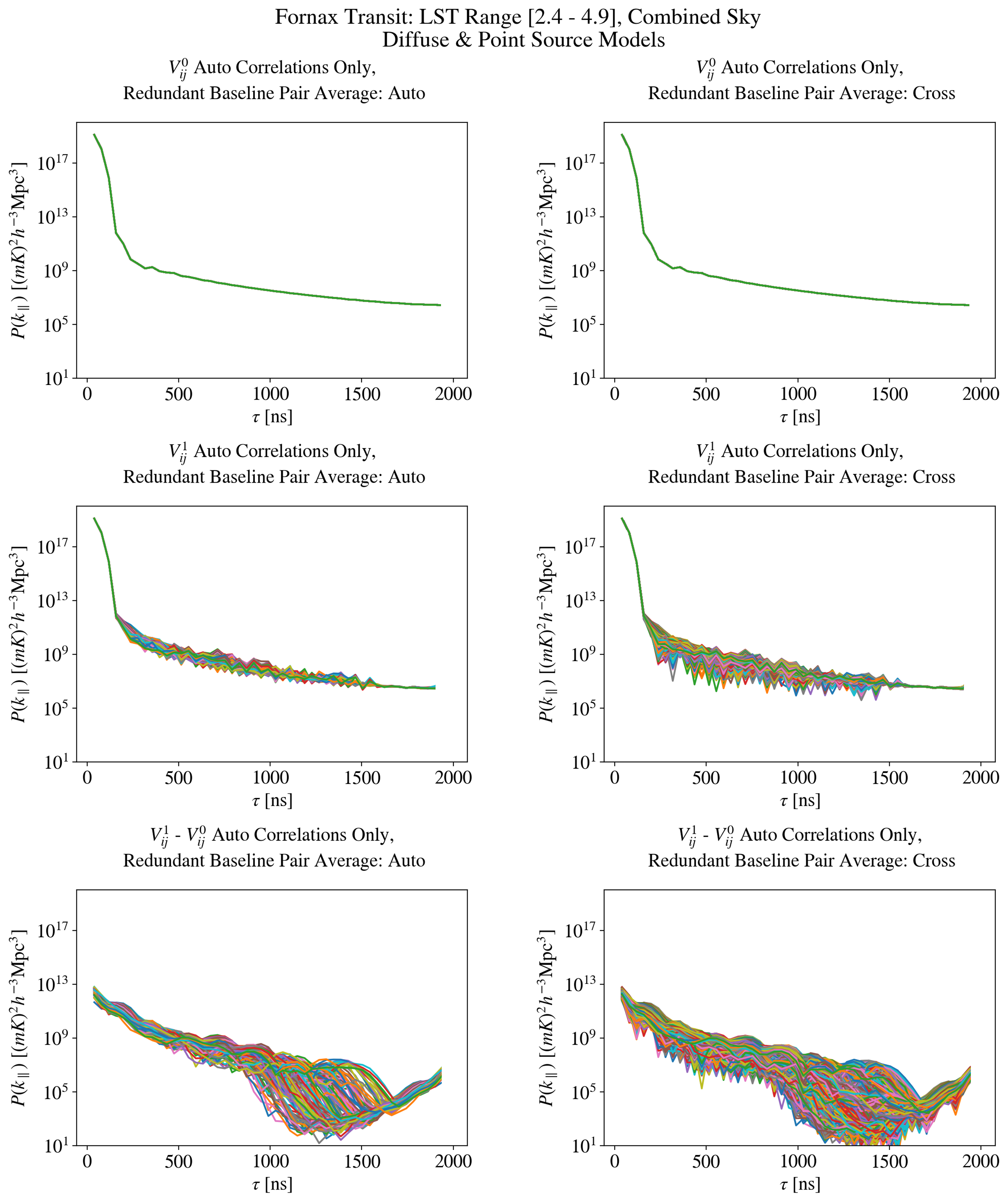}
    \caption{Time-averaged (but not baseline-averaged) power spectrum measurements of auto-correlation data, as a function of delay. The data products used in this plot are a pair of baselines, each belonging to the same redundant group. For example, this redundant group contain auto-correlations, e.g. ((176, 176, 'xx'), (176, 176, 'xx')) or ((162, 162, 'xx'), (178, 178, 'xx')). Data sets with an 'auto' baseline pair construction only contain baselines paired with itself [e.g. ((176, 176, 'xx'), (176, 176, 'xx'))]. Data sets with a 'cross' baseline pair construction contain baseline pairs with all other baseline pairs in the same redundant group, except for itself [e.g. ((162, 162, 'xx'), (178, 178, 'xx'))]. Frequency range: [144-169] MHz. LST range: [2.4-4.9], which captures Fornax A, a particularly bright point source, transiting the HERA beam. In this LST range, the HERA sky response at most (if not all) baselines is dominated by the point source model.}
    \label{fig:PSPEC_AUTO_BLP_FT}
\end{figure*}

Auto-correlations, which are (to zeroth-order) considered as being predominantly real-valued (i.e. having a negligible imaginary component), now contain (to first-order) a non-negligible imaginary component. Thus, first-order auto-correlation visibilities are no longer identical in the complex plane. In equation (\ref{eq:V_ij_0_matrix_simp}), aside from any window function associated with the beam term (which is discussed in Figure \ref{fig:delay_of_beam_window_func} and is typically of a low-enough delay to be negligible at zeroth-order), the only explicitly complex component of the visibility equation is the exponential $e^{2\pi i \frac{\nu}{c}\vec{b_{ij}}\cdot\hat{s}}$. For an auto-correlation baseline, this term is entirely real because $\vec{b_{ij}}=0$. However, first-order auto-correlations, per equation (\ref{eq:V_ij_1_with_0_simp}), have copies of cross-correlations (e.g. $V^0_{ki}$ which generally have non-zero imaginary components (because $\vec{b_{ki}}$ is not generally equal to zero). From Section (\ref{sec:PHENOMENON}), we know these copied visibilities will manifest themselves in different auto-correlations, $V^1_{ii}$, with different amplitudes depending on where in the array the specific antenna is located. Thus, the magnitude of auto visibilities is no longer real-valued and constant across the array: $imag(V^0_{ii}) \approx imag(V^0_{jj}) \approx 0; V^0_{ii} \approx V^0_{jj} \forall {i, j}$. Rather, first-order coupled auto-correlation visibilities are now complex-valued and unequal across the array: $imag(V^0_{ii}), imag(V^0_{jj}) \neq 0; V^0_{ii} \neq V^0_{jj}$. This non-redundancy is evident in a power spectrum measurement of auto-correlation data, regardless of whether we construct the power spectrum using auto baseline pairs or cross baseline pairs. Figures (\ref{fig:PSPEC_AUTO_BLP_GS}) and (\ref{fig:PSPEC_AUTO_BLP_FT}) compare the first and zeroth-order power spectra for the LST ranges of the galaxy setting in the horizon and Fornax A transiting the array, respectively. For the latter LST range, we find significantly less non-redundancy in the power spectrum.

\subsection{Contamination Outside of the Wedge: Power Spectrum Measurements as a Function of Baseline Length and Delay}
\label{sec:PSPEC_wedge}
\begin{figure*}
    \centering
    \includegraphics[width=0.99\textwidth]{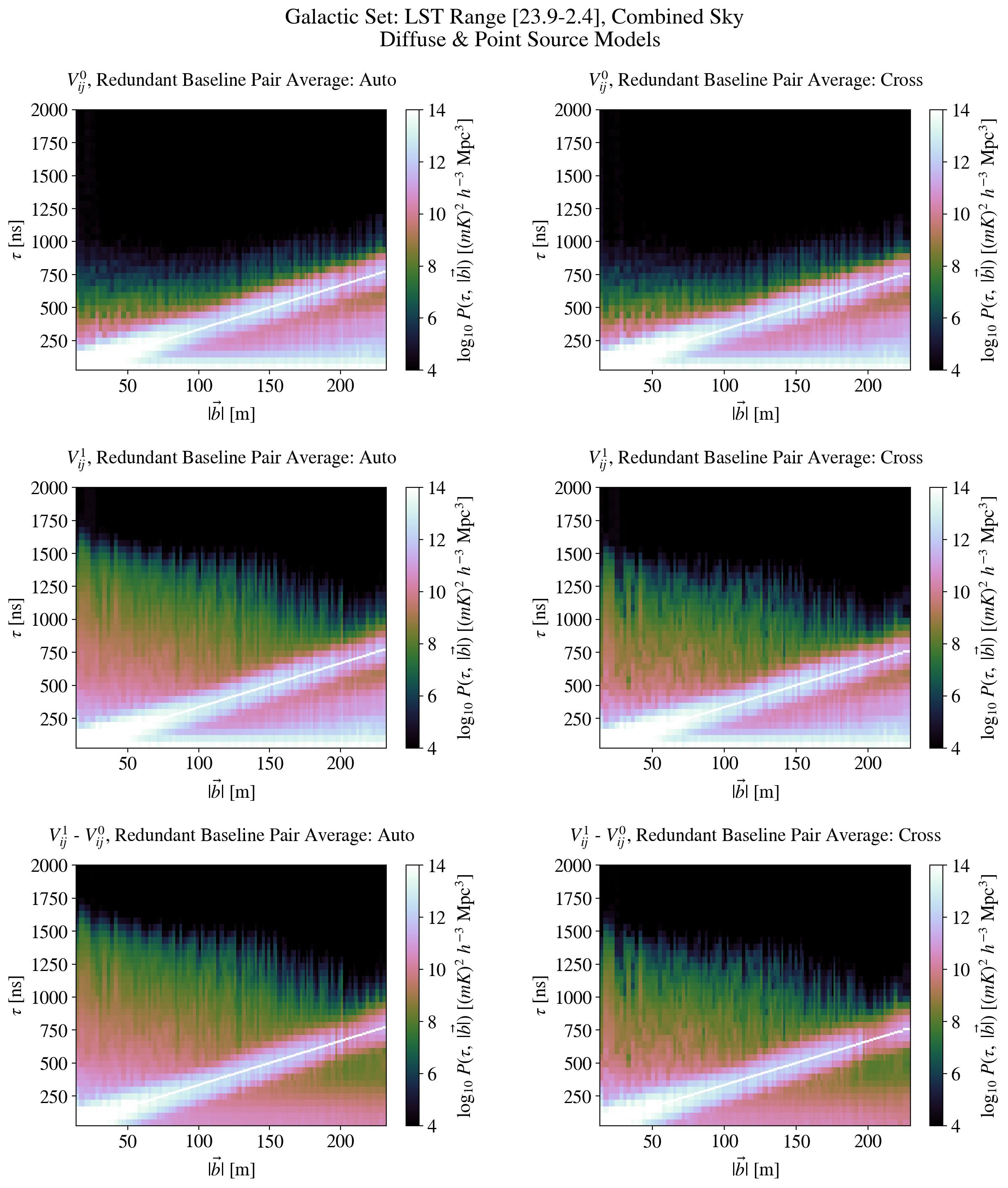}
    \caption{Time-averaged, baseline-averaged, power spectrum measurements as a function of baseline length and delay. The data products used in this plot are a pair of baselines, each belonging to the same redundant group. For example, the redundant group of 28m E/W baselines would contain the following pairs: ((176, 178, 'xx'), (176, 178, 'xx')) and ((162, 164, 'xx'), (176, 178, 'xx')). Data sets with a redundant baseline pair average of 'auto' are constructed by averaging all baseline pairs in a redundant group with only itself [e.g. ((176, 178, 'xx'), (176, 178, 'xx'))]. Data sets with a redundant baseline pair average of 'cross' are constructed by averaging a baseline pair with all other baseline pairs in the same redundant group, except for itself [e.g. ((162, 164, 'xx'), (176, 178, 'xx'))]. Frequency range: [144-169] MHz. LST range: [23.9-2.4], which corresponds to the galaxy setting in the horizon of HERA.}
    \label{fig:PSPEC_WEDGE_GS}
\end{figure*}

\begin{figure*}
    \centering
    \includegraphics[width=0.99\textwidth]{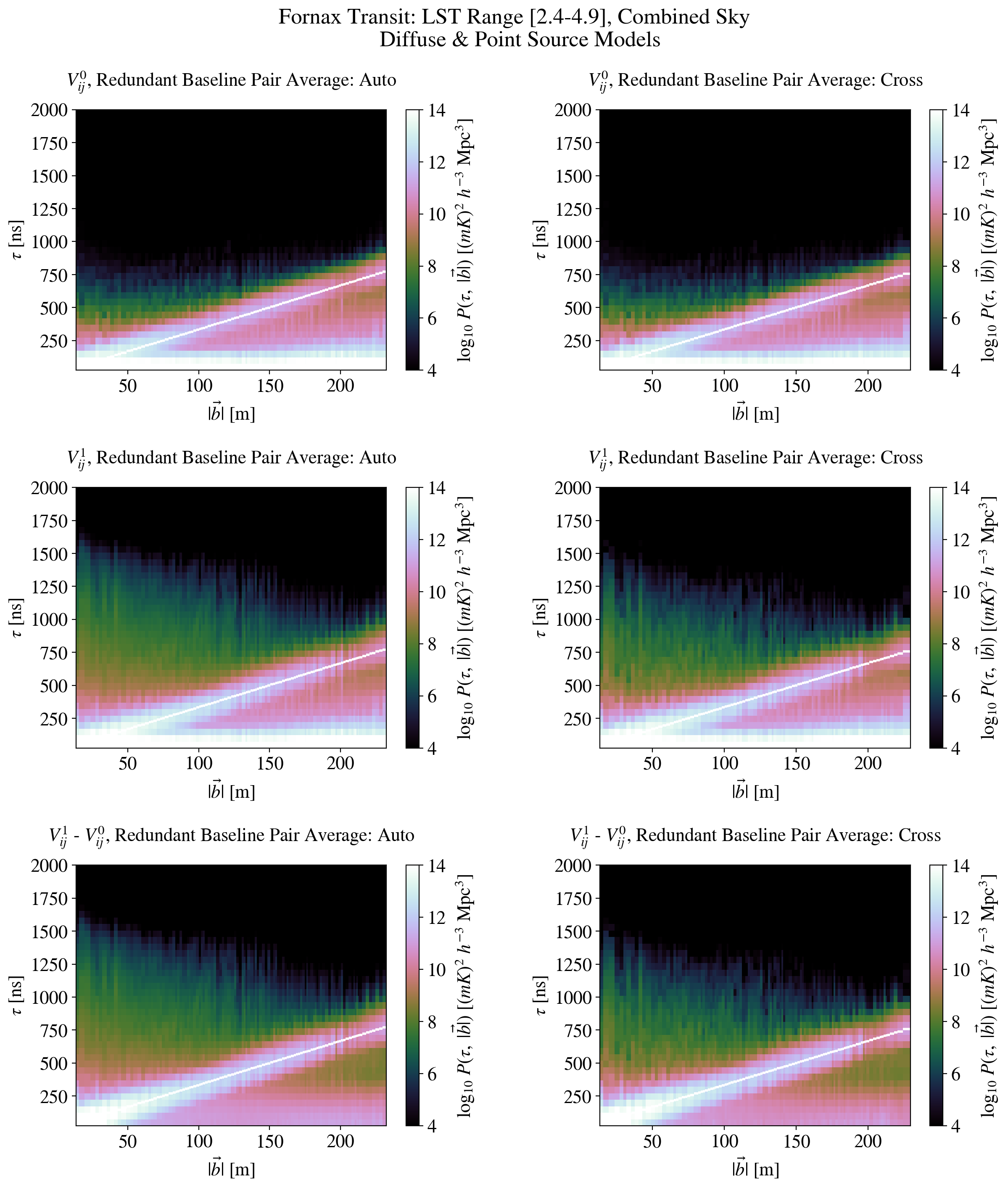}
    \caption{Time-averaged, baseline-averaged, power spectrum measurements as a function of baseline length and delay. The data products used in this plot are a pair of baselines, each belonging to the same redundant group. For example, the redundant group of 28m E/W baselines would contain the following pairs: ((176, 178, 'xx'), (176, 178, 'xx')) and ((162, 164, 'xx'), (176, 178, 'xx')). Data sets with a redundant baseline pair average of 'auto' are constructed by averaging all baseline pairs in a redundant group with only itself [e.g. ((176, 178, 'xx'), (176, 178, 'xx'))]. Data sets with a redundant baseline pair average of 'cross' are constructed by averaging a baseline pair with all other baseline pairs in the same redundant group, except for itself [e.g. ((162, 164, 'xx'), (176, 178, 'xx'))]. Frequency range: [144-169] MHz. LST range: [2.4-4.9], which captures Fornax A, a particularly bright point source, transiting the HERA beam. In this LST range, the HERA sky response at most (if not all) baselines is dominated by the point source model.}
    \label{fig:PSPEC_WEDGE_FT}
\end{figure*}

Figures (\ref{fig:PSPEC_WEDGE_GS}) and (\ref{fig:PSPEC_WEDGE_FT}) present power spectrum measurements for all possible redundant baseline groups in the array, at two different LST ranges, as a function of baseline length and delay. The top row of plots presents the "wedge" foreground structure, which has been studied extensively in the case of zeroth-order visibility, per the references cited in Section (\ref{sec:phenomenon_delay_contribs}). The middle rows of Figures (\ref{fig:PSPEC_WEDGE_GS}) and (\ref{fig:PSPEC_WEDGE_FT}) plot the same power spectrum measurement as the top row, but for first order visibilities. The bottom row of the figures presents the difference between the first and zeroth-order visibilities, which is equal to only the sum terms of equation (\ref{eq:V_ij_1_with_0_simp}). Equivalently, the bottom row of Figures (\ref{fig:PSPEC_WEDGE_GS}) and (\ref{fig:PSPEC_WEDGE_FT}) represents the power and structure uniquely associated with first-order coupling. 

\begin{figure*}
    \centering
    \includegraphics[width=0.99\textwidth]{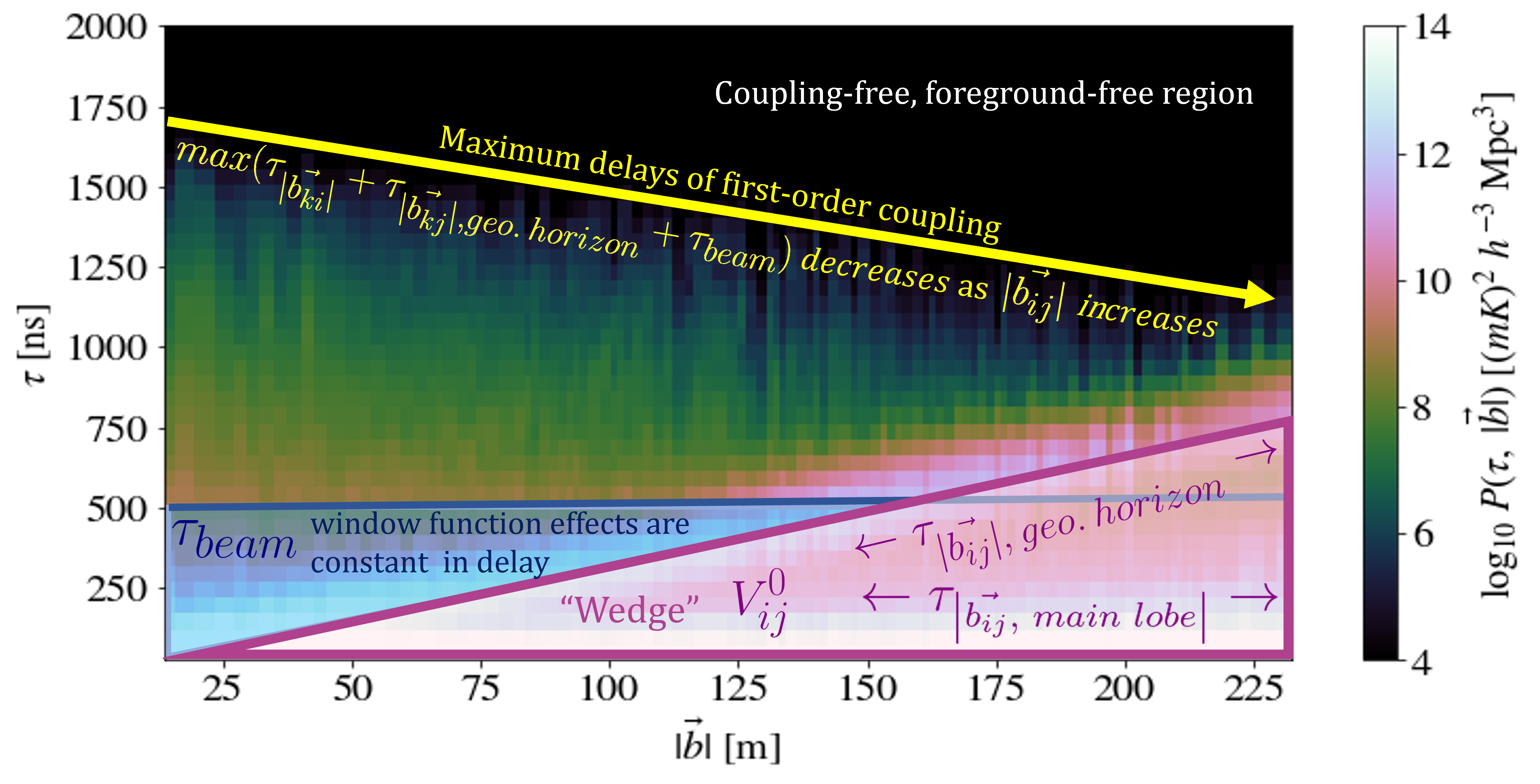}
    \caption{A conceptual outline of the different features of a wedge power spectrum containing first-order visibilities, as calculated using the simplified version of our coupling formalism, presented in equation (\ref{eq:V_ij_1_with_0_simp}). Section (\ref{sec:phenomenon_delay_contribs}) describes how first-order coupling features not only manifest themselves at delays corresponding to zeroth-order visibilities (e.g. $\tau_{|\vec{b_{ij}}|}$ and $\tau_{beam}$), but at much longer delays, e.g.  $max(\tau_{|\vec{b_{ki}}| } + \tau_{|\vec{b_{kj}}|, geo. \hspace{0.15cm} horizon} + \tau_{beam}  )$. This latter sum extends to delays beyond the 'wedge' which has been extensively studied in the literature and contains astrophysical foreground features in visibilities that do not consider first-order coupling. From equation (\ref{eq:V_ij_1_with_0_simp}), we note that first-order visibilities consist of copies of zeroth order visibilities ($V^0_{kj}$ and $V^0_{ik}$) which are translated to the delays presented in the last two cases of equation (\ref{eq:tau_coupling}). Super-wedge features associated with both the main lobe and pitchfork effect at the geometric horizon of the copied visibilities manifest themselves at delays which are inversely proportional to the copied baseline lengths, giving these first-order coupling features a distinct `inverse wedge' shape (yellow). For each baseline length, delays greater than the yellow line are free from first-order coupling effects. Figure (\ref{fig:ki_decrease_ij_increase}) offers insight as to why the relationship between baseline length and maximum delay of first-order coupling is inversely proportional. Figure (\ref{fig:delay_of_beam_window_func}) and Section (\ref{sec:phenomenon_delay_contribs}) describe the origin of $\tau_{beam}$.}
    \label{fig:PSPEC_WEDGE_DESCRIBED}
\end{figure*}

It is important to note that the maximum delay of the last two cases of equation (\ref{eq:tau_coupling}) can be greater than $\tau_{|\vec{b_{ij}}| }$. Thus, we expect first-order coupling to contaminate delay space beyond the wedge which contains zeroth-order foregrounds. This super-wedge contamination is clearly visible in Figures (\ref{fig:PSPEC_WEDGE_GS}) and (\ref{fig:PSPEC_WEDGE_FT}), and is conceptually documented in Figure (\ref{fig:PSPEC_WEDGE_DESCRIBED}). The "inverse-wedge" feature presented in these figures depict the maximum values of the last two delay contributions presented in equation (\ref{eq:tau_coupling}). Figure (\ref{fig:ki_decrease_ij_increase}) is a visual aid for explaining how the super-wedge features of equation (\ref{eq:tau_coupling}) can actually appear at such high delays. Such wedge contamination, if not mitigated, may pose a critical impasse for a HERA-like interferometer to make a detection of the EoR. 

When comparing the first column to the second column in Figure (\ref{fig:PSPEC_WEDGE_GS}), or also making the same comparison in (\ref{fig:PSPEC_WEDGE_FT}), we note that the lower-delay 'inverse wedge' feature associated with $max(\tau_{|\vec{b_{ki}}| } + \tau_{V^{0}_{kj, main \hspace{0.1cm} lobe} } + \tau_{beam}  )$, reduces in power when averaging all cross baseline pairs, compared to only averaging auto baseline pairs. This is expected behavior, because intuitively we consider different baselines to receive incident astrophysical radiation with different phases from one another, even if all baselines belong to the same redundant group, by the fact that the baselines are physically located at different points in the array. 
It is also intuitive that the coupling features in Figure (\ref{fig:PSPEC_WEDGE_GS}) are brighter than those in Figure (\ref{fig:PSPEC_WEDGE_FT}) because the apparent brightness of astrophysical radiation at the horizon is brighter in the LST range associated with the setting of the galaxy than it is in the LST range associated with the transit of Fornax A; see Figure (\ref{fig:beam_apparent_GS_FT}). 

\subsection{Delay-Based Coupling Mitigation Strategy}
\label{sec:pspec_delay_mitigration}
There exist subsets of our analysis space where first-order power spectra are consistent with their zeroth-order counterparts down to at least 10 orders of magnitude in sensitivity. In Figures (\ref{fig:PSPEC_WEDGE_GS}) and (\ref{fig:PSPEC_WEDGE_FT}), this region exists for each baseline group at delays greater than the maximum extent of case 4 of equation (\ref{eq:tau_coupling}); this foreground-free, coupling-free region is denoted in Figure (\ref{fig:PSPEC_WEDGE_DESCRIBED}) as delays greater than the yellow line. This region could be critical to any future coupling mitigation strategy, although further analysis of simulated data, and comparing said data to observation, is required to confirm  whether such a strategy will mitigate first-order coupling effects to levels below the expected power of the 21cm EoR signal.

\section{Discussions}\label{sec:DISCUSSIONS}
\subsection{The Strengths and Weakness of a Semi-Analytic Coupling Model, Motivating the Use of Embedded Element Patterns}
The steady-state formalism described above considers the effect of first-order antenna-antenna coupling in interferometric visibilities. Incident astrophysical radiation enters all interferometric elements in the array. Each element absorbs most of its incident radiation, although some amount of power is then reflected due to departure from conjugate match at the terminals of the antenna feeds. Subsequently, each element in the array re-radiates power across the array. By being a first-order formalism, we only model this coupling effect on visibilities when re-radiated power is absorbed by just one more element, not when the transmitted power is subsequently absorbed and re-radiated by other elements. Such a model produces coupling systematics which extend beyond the wedge in baseline length versus delay space which is thought to contain most zeroth-order astrophysical foregrounds, at both zero and nonzero fringe rates.

The benefit of limiting our analysis to first-order antenna-antenna coupling is that we are able to straightforwardly produce a closed-form expression for coupled visibilities, e.g. equation (\ref{eq:V_ij_1_with_0_matrix}). Upon deriving a closed-form expression, Section \ref{sec:sim_0} of this work lays the framework for a simulation of zeroth-order visibilities which are coupled to first order. 

While our first-order phenomenological analysis (Section \ref{sec:PHENOMENON}) benefits from the fact that it can easily be simulated, we warn the reader that first-order antenna-antenna coupling effects cannot generally be considered the dominant mutual coupling effect. For example, note that the coupling terms in (\ref{eq:V_ij_1_with_0_matrix}) are attenuated by baseline lengths ($|\hat{b}_{ki}|$ and $|\hat{b}_{kj}|$) and voltage reflection coefficients ($\Gamma_k$ $<0$, $\forall$k). Depending on how conjugate mismatched a feed is to its terminals (essentially, by comparing the ratio of $\Gamma_k$ to $|\hat{b}_{ki}|$ and $|\hat{b}_{kj}|$), a second-order coupling event of a short baseline could be stronger in magnitude than a first-order coupling event from a longer baseline. It is beyond the scope of this document to analyze how common a HERA-like array experiences second-order coupling effects which are stronger in magnitude than the first-order effects derived above. 

To account for all orders of antenna-antenna coupling in the general visibility formalism, one could calculate the Active Array Element Pattern (also referred to as the Embedded Element Pattern) of each element in the array. This data product, originally considered in \cite{1138237}, is analogous to the "effective height" terms in equations (\ref{eq:v_i_0}) and (\ref{eq:v_j_0}). In an active array element pattern, \emph{all} orders of antenna coupling (not just first-order coupling) are considered by modifying the gain of an array element as a function of azimuth, zenith, and frequency. Assuming we had the computational resources to calculate these embedded element patterns (whether using commercial electromagnetic simulation software, such as CST, or by means of a proprietary software), the sums present in equations (\ref{eq:v_i_1}) and (\ref{eq:v_j_1}) would not need to be explicitly considered, as they would be assumed into the embedded element pattern. In essence, the embedded element pattern would replace $h^{px}_{i}(\hat{s})$ and $h^{qy}_{j}(\hat{s})$. It is beyond the scope of this work to derive a general visibility formalism which uses embedded element patterns, although we note that the concept of using embedded element patterns has previously been applied in other radio interferometric applications, such as \cite{1705.08109}.

\subsection{Mutual Coupling: A Thread to Redundant Baseline Averaging and to Wedge Foreground Containment}
Comparing Figures (\ref{fig:176_178_GS}) and (\ref{fig:176_178_FT}) to other baselines of the respective LSTs, which are found in the online supplementary material, it is clear that baselines with the same ENU coordinates experience first-order coupling differently; this is a function of where each specific baseline is located in the array. This non-redundancy in a baseline group is summarized in the power spectrum measurements of Figures (\ref{fig:PSPEC_AUTO_BLP_GS}) and (\ref{fig:PSPEC_AUTO_BLP_FT}). Redundant arrays such as HERA (or its pathfinder project, PAPER), were commissioned with the assumption that data from baseline groups with identical ENU coordinates could be averaged together. Any introduction of non-redundancy in such a baseline group threatens the observation strategy, power spectrum sensitivity requirements, and overall analysis pipeline of such an array. 

Furthermore, the general analysis approach of interferometric arrays has focused on discriminating the EoR signal from spectrally smooth astrophysical foregrounds which are many orders of magnitude brighter. One prevailing, and reasonably conservative strategy for such foreground mitigation, discussed primarily in \citet{Parsons2012} and \citet{WEDGE_NITHYA_2015} is 'avoidance': One seeks to measure the EoR at delays and interferometric baseline lengths which are entirely distinct from the delays and baseline lengths associated with foregrounds. Such strategies discuss a 'wedge' of foreground space, such as those presented in the top rows of Figures (\ref{fig:PSPEC_WEDGE_GS}) and (\ref{fig:PSPEC_WEDGE_FT}). The ability to entirely avoid foreground contamination is threatened by first-order coupling features which extend beyond the zeroth-order wedge, and are conceptually described in Figure (\ref{fig:PSPEC_WEDGE_DESCRIBED}). Other analyses, such as \citet{Kern:2019}, discuss filtering strategies for non-fringing coupling features, but such strategies are not sufficient for mitigating coupling features with non-zero fringe, such as those characterized in this work. 

\subsection{Array Designs and Observation Strategies which Mitigate Coupling Effects}
First-order array element coupling effects are a function of LST, baseline length, and baseline orientation. Since it is a a function of all three effects, a general strategy for filtering mutual coupling is not trivial and beyond the scope of this analysis. Nevertheless, the presented analysis motivates how our semi-analytic coupling model may be used by the reader to mitigate coupling systematics in existing radio interferometers and to design future arrays where the configuration of array elements inherently mitigates coupling effects at desired LSTs and/or angular resolutions on the sky.

While it is beyond the scope of this work to fully discuss how one could position the elements of an interferometric array in order to mitigate fringing and non-fringing coupling effects, we briefly outline a mitigation strategy for the observation of only one particular LST range. Suppose, for example, that an experiment requires an interferometric array to only observe the galactic center. In this LST range, the apparent brightness of the sky in any array element would be at zenith. By (\ref{eq:FRINGERATE}), the principle first-order coupling features will be concentrated at fringe rates of $f_r(0, 0) \approx \frac{b\nu}{c} \cos \alpha \cos \delta$, which is proportional to a baseline's projected E/W length ($b \cos \alpha$). At this desired LST range, coupling effects would be entirely avoided if all baselines used in power spectrum measurements were configured in a N/S orientation (e.g. row 4, column 3 of Figure (\ref{fig:176_178_FT}) and Figures (S2), (S6), and (S8) of the online supplemental material).

Similarly, an observer could mitigate first-order coupling effects at a particular angular resolution on the sky (i.e. for one or a few interferometric baseline lengths) if he chose to only make measurements at LSTs where the apparent brightness of the sky in any array element were at zenith.

\section{Conclusion}\label{sec:CONCLUSION}
We derive a general, fully-polarized, semi-analytic formalism for interferometric visibilities, which considers how impedance mismatch at the feed terminals of each element in an array will cause incident astrophysical radiation to re-radiate and couple into all other elements in the array. Our model assumes steady-state, incident radiation. Since it is a first-order model, we only capture the effect of re-radiation being absorbed by all other elements in the array just once (e.g. sky $\rightarrow$ Antenna k $\rightarrow$ Antenna i), not when re-radiated power is subsequently absorbed and re-radiated to all elements multiple times (e.g. sky $\rightarrow$ Antenna k $\rightarrow$ Antenna i $\rightarrow$ Antenna j).  We simulate first-order coupling features on a HERA-like redundant array using non-polarized skies with diffuse and point-source emission; a phenomenological analysis of the coupling features is presented. Contrary to previous studies, we find mutual coupling features manifest themselves at nonzero fringe rates. We compare power spectrum results for both coupled and non-coupled (noiseless, simulated) data and find coupling effects to be highly dependent on LST, baseline length, and baseline orientation. For all LSTs, lengths, and orientations, coupling features appear at delays which are outside the foreground 'wedge' containing non-coupled astrophysical foreground features. Further, we find that first-order coupling effects threaten our ability to average data from baselines with identical length an orientation. We propose two filtering strategies, one in delay space and the other in fringe-rate space, which may mitigate such coupling systematics. The semi-analytic coupling model herein presented may be used to study mutual coupling systematics as a function of LST, baseline length, and baseline orientation. Such a model is not only helpful to the field of 21cm cosmology, but any study involving interferometric measurements, where coupling effects at the level of at least 1 part in $10^4$ could corrupt the scientific result. Our model may be used to mitigate coupling systematics in existing radio interferometers and to design future arrays where the configuration of array elements inherently mitigates coupling effects at desired LSTs and angular resolutions. 

\section*{Acknowledgements}
We thank James Aguirre, Jacqueline Hewitt, Nicholas Kern, Daniel Jacobs, and Vincent MacKay for discussions which directly contributed to this work. Our compliments to Bang Nhan and Honggeun Kim for the high-precision re-simulation of Nicolas Fagnoni's CST models of the HERA Vivaldi feed, which were used to construct all beam products analyzed in this paper. This material is based upon work supported by the National Science Foundation under Grant Nos. 1636646 and 1836019 and institutional support from the HERA collaboration partners. This research is funded in part by the Gordon and Betty Moore Foundation. HERA is hosted by the South African Radio Astronomy Observatory, which is a facility of the National Research Foundation, an agency of the Department of Science and Innovation.
\section*{Software and Data Availability}
A demonstration of how to apply the first-order coupling formalism to interferometric visibility data may be found in the first author's public Github repository \footnote{https$://$github.com$/$alphatangojuliett$/$CoupledRadioInterferometer}.The healvis visibility simulator may be found in \citet{2019ascl.soft07002L}\footnote{https$://$github.com$/$rasg-affiliates/healvis}. Simulated visibilities (both zeroth-order and first-order) are output as uvh5 data objects, which may be conveniently read and accessed via pyuvdata (\citet{Hazelton2017})\footnote{https$://$github.com/RadioAstronomySoftwareGroup/pyuvdata}. Running this analysis on simulated HERA-like data products was made substantially more convenient using the 'i/o' functions and data containers defined in the hera$\_$cal repository \footnote{https$://$github.com/HERA-Team/hera$\_$cal}. We use the hera$\_$pspec power spectrum estimator to calculate all power spectra herein presented \footnote{https$://$github.com/HERA-Team/hera$\_$pspec}. Additional plots of coupled and non-coupled visibilities, as a function of fringe-rate versus delay, may be found in the the online supplemental material. Simulated data of the electromagnetic properties of HERA, including CST beam products and reflection coefficients, may be requested via email from the authors.
\bibliographystyle{mnras}
\bibliography{main} 


\appendix
\section{Motivating the Standard Visibility Equation}
\label{sec:general_formalism_motivation} 
\begin{figure}
    \centering
    \includegraphics[width=.26\textwidth]{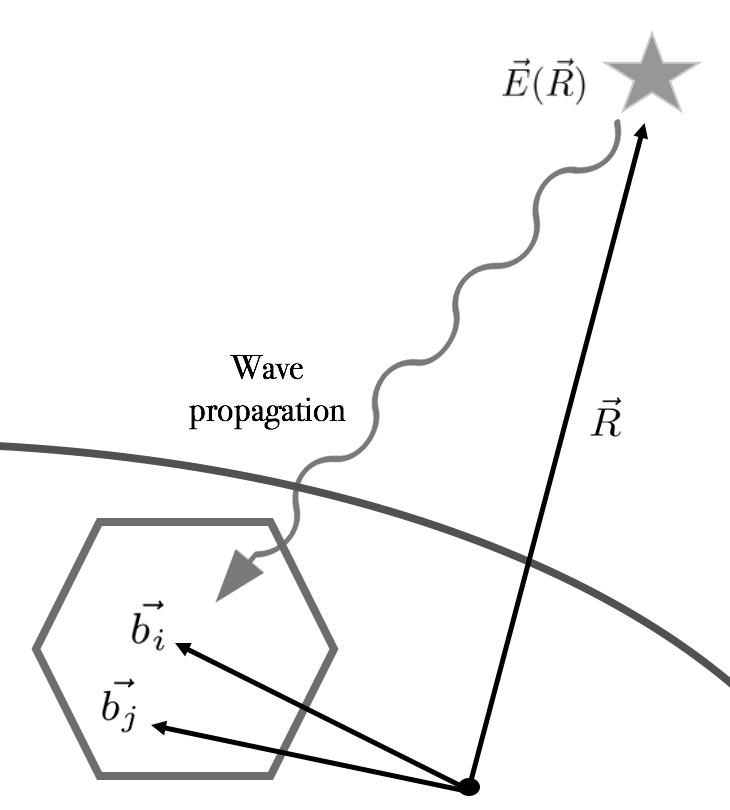}
    \caption{Electric field propagation from astrophysical source to two elements of an interferometric array. Both the source and the interferometric elements in the array are referenced to an arbitrary ground point.  }
    \label{fig:EM_propagation}
\end{figure}

Before establishing a visibility equation, we begin by discussing the propagation of radiation from a single astrophysical source to an interferometric array on Earth.  By the linearity of Maxwell's equations, we may superimpose the electric fields at all measurement locations, $b_i$, by the source point. The propagation of this electric field from the source to a test point can be characterized by some function $P(\vec{R}, \vec{b_i})$.
\begin{equation}\label{eq:E_Maxwell}
\int \int \int P(\vec{R}, \vec{b_i}) \vec{E}(\vec{R})dxdydz
\end{equation}

We now make our first assumption, Assumption A, which is both logical and also serves as the basis for standard visibility derivations, such as \citet{1999ASPC..180....1C}. Namely, we assume that the astrophysical sources are very far away from the array, so that $|\vec{R}| >> |\vec{b_i}|$. In this case, all we can hope to characterize is the surface brightness of a source, and information about brightness along the radial axis is lost. We imagine a 'celestial sphere' of radius $|\vec{R}|$, and focus on integrating the electric field surface distribution, $\vec{\varepsilon}(\vec{R})$.

We now introduce Assumption B, that the space within the celestial sphere is empty. Thus, by Huygen's Principle for spherical wave propagation, the propagation function of (\ref{eq:E_Maxwell}) is
\begin{equation}\label{eq:Propagation}
P(\vec{R}, \vec{b_i}) = e^{ 2\pi i \frac{\nu}{c}|\vec{R}-\vec{b_i}| }
\end{equation}
Thus, the electric field at interferometric element $i$, as measured by integrating over surface area elements $ds$, is
\begin{equation}\label{eq:E_at_i}
\vec{E}(\vec{R}, \vec{b_i}) = \int \vec{\varepsilon}(\vec{R}) \frac{e^{ 2\pi i \frac{\nu}{c}|\vec{R}-\vec{b_i}| }}{|\vec{R}-\vec{b_i}|}ds
\end{equation}

We now imagine that each interferometric element consists of an electromagnetically sensitive structure, such as a feed, or a dish with a feed near the dish's optical focus. The structure has two orthogonal feed polarizations, $p$ and $q$. Assuming that the electric field of the astrophysical sources propagates at a frequency which is in the electromagnetically sensitive band of each feed polarization, then the open-circuit voltage of the $p$-oriented feed of the $i^{th}$ element is
\begin{equation}\label{eq:V_eq_h_E}
v^p_i = \vec{h}^p_i \cdot \vec{E}
\end{equation}

Let us represent our array's interferometric elements in terms of local, Cartesian, ENU coordinates, $(x,y,z)$ and the electric field components in terms of orthogonal spherical coordinates $(r, \theta, \phi)$, where (in radians) $0 \leq \theta \leq \pi$ and  $0 \leq \phi \leq 2\pi$. Recall that $\hat{\theta}$ and $\hat{\phi}$ are tangent to the celestial sphere. Figure (\ref{fig:Spherical_Coordinates}) graphically represents this system, whose relationship to Cartesian coordinates is
\begin{align}\label{eq:r_hat_spherical}
& \hat{r} = \cos(\phi) \sin(\theta) \hat{x} +  \sin(\phi) \sin(\theta) \hat{y} + \cos(\theta) \hat{z} \\
& \hat{\theta} = \cos(\phi) \cos(\theta) \hat{x} +  \sin(\phi) \cos(\theta) \hat{y} - \sin(\theta) \hat{z} \\
& \hat{\phi} = -\sin(\phi) \hat{x} +  \cos(\phi) \hat{y}
\end{align}
Noting that our differential surface area element $ds$ points in the radial direction, $\hat{s}=\frac{\vec{R}}{|\vec{R}|}$, we can integrate over a differential solid angle, $d\Omega$, where $ds = R^2 d\Omega$. Thus, the voltage response at the terminal of the $p^{th}$ pol of the $i^{th}$ interferometric element, caused by the surface distribution of the $\hat{\phi}$ basis vector component of the electric field is

\begin{equation}\label{eq:v_i_0_DERIVED}
v^{p\phi}_{i} = \int h^{p\phi}_{i}(\hat{s})  \varepsilon^{\phi}(\hat{s})\frac{e^{2\pi i \frac{\nu}{c}|R\hat{s}-\vec{b_i}| }}{|R\hat{s}-\vec{b_i}|}R^2 d\Omega
\end{equation}

Similarly, the voltage response at the terminal of the $q^{th}$ pol of the $j^{th}$ interferometric element, caused by the surface distribution of the $\hat{\theta}$ basis vector component of the electric field is
\begin{equation}\label{eq:v_j_0_DERIVED}
v^{q\theta}_{j} = \int h^{q\theta}_{j}(\hat{s}^{'}) \varepsilon^{\theta}(\hat{s}^{'})\frac{e^{2\pi i \frac{\nu}{c}|R\hat{s}^{'}-\vec{b_j}| }}{|R\hat{s}^{'}-\vec{b_j}|}R^2 d\Omega^{'}
\end{equation}

The above voltage responses do not consider reflections of incident radiation due to impedance mismatch at each feed. This work considers 1st-order re-radiation of the astrophysical electric field, which arrives at the $i^{th}$ and $j^{th}$ elements after being previously re-radiated by only one other element $k$ (and not more than one element). Because the array may be located generally on the geographic (2D) plane, not simply in a (1D) line, the scattered electric field (from the $k^{th}$ element), upon arrival to any instrumental polarization of all other elements, can be a function of $\emph{both}$ instrumental polarizations of the radiating ($k^{th}$) element. Consequently, the scattered electric field can be a function of both incident astrophysical electric field basis vectors. We may generally define the electric field surface distribution which is re-radiated by a $k^{th}$ element, as received by the $i^{th}$ element, as $\varepsilon^{s\phi,p\theta}(\hat{s}, \hat{b_{ki}})$. The first superscript (e.g. '$\phi$') refers to $\hat{\phi}$ component of what is received by the $i^{th}$ element from the radiating $k^{th}$ element. After the comma in the superscript (e.g. $p\theta$), we denote the incident electric field basis vector (e.g. $\hat{\theta}$) contribution to each instrument polarization (e.g. $p$) of the radiating $k^{th}$ element. A similar, general expression is used for the electric field distribution which is scattered from the $k^{th}$ element to the $j^{th}$ element, e.g. $\varepsilon^{s\theta,q\phi}(\hat{s}, \hat{b_{kj}})$.

To summarize, the voltage response at the terminal of the $p^{th}$ pol of the $i^{th}$ interferometric element, caused by the $\hat{\phi}$ basis vector component of the electric field distribution which was re-radiated from the $k^{th}$ element may contain components from both instrumental polarizations and both incident electric field basis vectors of the $k^{th}$ element. Similarly, the voltage response at the terminal of the $q^{th}$ pol of the $j^{th}$ interferometric element, caused by the $\hat{\theta}$ basis vector component of the electric field distribution which was re-radiated from the $k^{th}$ element may contain components from both instrumental polarizations and both incident electric field basis vectors of the $k^{th}$ element. We formalize this model of antenna-to-antenna coupling by re-writing equations (\ref{eq:v_i_0_DERIVED}) and (\ref{eq:v_j_0_DERIVED}):

\begin{align}\label{eq:v_i_1}
    v^{p\phi}_{i} & = \int h^{p\phi}_{i}(\hat{s}) \varepsilon^{\phi}(\hat{s})\frac{e^{2\pi i \frac{\nu}{c}|R\hat{s}-\vec{b_i}| }}{|R\hat{s}-\vec{b_i}|}R^2 d\Omega \\
    & +\sum_{k \neq i}\int \frac{e^{2\pi i \frac{\nu}{c}|R\hat{s}-\vec{b_k}|}}{|R\hat{s}-\vec{b_k}|} \bigg[ \varepsilon^{s\phi,p\phi}(\hat{s}^{'}, \hat{b_{ki}})+\varepsilon^{s\phi,p\theta}(\hat{s}^{'}, \hat{b_{ki}}) \nonumber \\ 
    & +\varepsilon^{s\theta,q\phi}(\hat{s}^{'}, \hat{b_{ki}})+\varepsilon^{s\theta,q\theta}(\hat{s}^{'}, \hat{b_{ki}}) \bigg] h^{p\phi}_{i}(\hat{b_{ik}}) R^2 d\Omega \nonumber 
\end{align}

\begin{align}\label{eq:v_j_1}
    v^{q\theta}_{j} & = \int  h^{q\theta}_{j}(\hat{s}^{``}) \varepsilon^{\theta}(\hat{s}^{``})\frac{e^{2\pi i \frac{\nu}{c}|R\hat{s}^{``}-\vec{b_j}| }}{|R\hat{s}^{``}-\vec{b_j}|}R^2 d\Omega^{``} \\ 
    & + \sum_{k \neq j}\int  \frac{e^{2\pi i \frac{\nu}{c}|R\hat{s}^{``}-\vec{b_k}| }}{|R\hat{s}^{``}-\vec{b_k}|} \bigg[ \varepsilon^{s\phi,p\phi}(\hat{s}^{'}, \hat{s}^{'}_{kj})+\varepsilon^{s\phi,p\theta}(\hat{s}^{'}, \hat{s}^{'}_{kj}) \nonumber \\
    & +\varepsilon^{s\theta,q\phi}(\hat{s}^{'}, \hat{s}^{'}_{kj})+\varepsilon^{s\theta,q\theta}(\hat{s}^{'}, \hat{s}^{'}_{kj}) \bigg] h^{q\theta}_{j}(\hat{s}^{``}_{jk}) R^2 d\Omega^{``} \nonumber
\end{align}

\section{Compatibility of HERA Beam with Assumptions in the General Formalism}
\label{sec:BEAM_COMPATIBILITY}
\subsection{Satisfaction of Far-Field Limit in HERA's Horizon Plane}
\label{sec:sim_far_field}
To simulate antenna-to-antenna scattering in a HERA-like array using the general formalism, we must first establish that the far-field approximation used in Equation (\ref{eq:E_scatter_correct_propagator}) is valid. In HERA, each of the Vivaldi feeds (which contain the feed terminals) are positioned above the rims of each dish. HERA is a drift-scan array with the peak beam gain of all feeds oriented directly towards zenith. Conventionally, one would considering the far field limit of radiation coming from zenith. This is guaranteed for incident astrophysical radiation by Assumption A. However, in our coupling formalism, the scattered radiation comes exclusively from the horizon plane (e.g. $\hat{b_{ik}}$, $\hat{b_{kj}}$). Thus, we must ensure that far field limit is not only satisfied for incident astrophysical radiation (coming from zenith), but is also satisfied for scattered radiation along the horizon direction, emitted at distances as short as $14m$. It can be shown by analyzing CST electromagnetic simulations of the beam that, in the horizon plane, the beam gain of a Vivaldi feed with and without the surrounding dish are nearly identical. Our result implies that, in the horizon plane (but not in the zenith direction), the feed alone will be the dominant radiator of a scattered electric field due to impedance mismatch at the feed terminals. 

Let us now ensure that this feed-to-feed radiation satisfies the far-field limit. We consider the far-field limit to be valid at distances greater than the Fraunhofer distance, $d_{F}$, where 
\begin{equation}\label{eq:d_Fraunhofer}
d_{F} = \frac{2D^2}{\lambda}
\end{equation}
HERA's Vivaldi feeds have a diameter approximately equal to $D\approx2$m. HERA operates in the range 50-250 MHz, corresponding to $\lambda \in [1.2, 6]$m, separated by a minimum of $14$m. Thus, the far-field limit is practically satisfied for the re-radiated electric field of all $\hat{b_{ki}}$ baseline directions considered in (\ref{eq:V_ij_1_with_0_matrix}), allowing us to apply the general formalism to HERA. 

\subsection{Relating EM Simulation (CST) Outputs to Formalism}
\label{sec:sim_simple_CST}

Using the definition of our Jones matrix in terms of effective height (\ref{eq:Jones_Matrix_Relationship}), we calculate
\begin{equation}\label{eq:J_Jdagger_Matrix}
\textbf{J}(\hat{s})\textbf{J}^{\dagger}(\hat{s})= 
\begin{bmatrix}
|h^\phi_p(\hat{s})|^2 + |h^\theta_p(\hat{s})|^2  & h^\phi_p h^{*\phi}_q(\hat{s}) + h^\theta_p h^{*\theta}_q(\hat{s})\\
h^\phi_q h^{*\phi}_p(\hat{s}) + h^\theta_q h^{*\theta}_p(\hat{s}) & |h^\phi_q(\hat{s})|^2 + |h^\theta_q(\hat{s})|^2 
\end{bmatrix}.
\end{equation}
Typically, this beam term is normalized so that the power at some specific, peak point is 1.  For example, to normalize such that the power (per polarization $p$) at zenith is 1, we stipulate that $1=|\textbf{J}_p(\hat{s_z})|^2$, where $\hat{s_z}$ is the unit vector pointing in the zenith direction. When we simulate the beam of a HERA interferometric element in CST (whether it be just the Vivaldi feed, or we also include the dish) there are two primary data products: First, a unitless directivity beam, $D_p(\hat{s})$, which may similarly be normalized at zenith. Second, an electric field beam, $\textbf{E}(\hat{s})$, to be discussed below. Both beams use the spherical coordinate basis vectors represented in Figure (\ref{fig:Spherical_Coordinates}).

The coupling terms in the first-order visibility formalism contain an additional multiplication of Jones matrices, this time pointing in the direction of the baseline being considered. Namely, $\textbf{J}(\hat{b_{ik}})\textbf{J}(\hat{b_{ki}})$ and $\textbf{J}^{\dagger}(\hat{b_{kj}})\textbf{J}^{\dagger}(\hat{b_{jk}})$. These terms do not simplify nicely into power beams, because they are not conjugate pairs. Rather, we have the multiplication of two complex numbers where neither of the numbers, or both of the numbers, are conjugated. Physically, we interpret this behaviour as re-radiation due to the antenna impedance at the feed terminals not being conjugate-matched to the rest of the analog chain. Mathematically, this is why we see "double-dagger" or "no-dagger" terms multiplied by $\Gamma$ (the voltage reflection coefficient) in (\ref{eq:V_ij_1_with_0_matrix}). 

It can be shown using orthographic projections of the HERA beams that, in the CST electric field basis vector convention, the $\hat{\phi}$ component of the electric field remains unchanged regardless of whether the baseline vector points towards the radiating or receiving element, 
$\textbf{J}^{\phi}(\hat{b_{ik}}) = \textbf{J}^{\phi}(\hat{b_{ki}})$. It may similarly be shown that, when the baseline vectors reside in the horizon plane ($\theta \approx \frac{\pi}{2}$), the $\hat{\theta}$ component of the beam flips sign:  $\textbf{J}^{\theta\approx\frac{\pi}{2}}(\hat{b_{ik}}) \approx -\textbf{J}^{\theta\approx\frac{\pi}{2}}(\hat{b_{ki}})$. Incorporating this sign flip for the $\hat{\theta}$ basis vector components, the Jones matrices in the coupling terms may be explicitly written as
{\small
\begin{align}\label{eq:JJ_Matrix}
\textbf{J}(\hat{b_{ik}})\textbf{J}(\hat{b_{ki}}) = \nonumber \\
\begin{bmatrix}
h^\phi_p h^{\phi}_p(\hat{b_{ki}}) - h^{\phi}_qh^\theta_p(\hat{b_{ki}})  & h^\theta_p h^{\theta}_q(\hat{b_{ki}}) - h^{\theta}_p h^\phi_p(\hat{b_{ki}}) \\
h^{\phi}_ph^\phi_q(\hat{b_{ki}}) - h^\phi_q h^{\theta}_q(\hat{b_{ki}}) & h^\theta_q h^{\theta}_q(\hat{b_{ki}}) - h^{\theta}_p h^\phi_q(\hat{b_{ki}})
\end{bmatrix} 
\end{align}

\begin{align}\label{eq:Jdagger_Jdagger_Matrix}
    \textbf{J}^{\dagger}(\hat{b_{kj}})\textbf{J}^{\dagger}(\hat{b_{jk}}) = \nonumber \\
    \begin{bmatrix}
    h^{*\phi}_p h^{*\phi}_p(\hat{b_{kj}}) - h^{*\phi}_qh^{*\theta}_p(\hat{b_{kj}})  & h^{*\phi}_qh^{*\phi}_p(\hat{b_{kj}}) - h^{*\phi}_q h^{*\theta}_q(\hat{b_{kj}}) \\
    h^{*\theta}_ph^{*\theta}_q(\hat{b_{kj}}) - h^{*\theta}_p h^{*\phi}_p(\hat{b_{kj}}) & h^{*\theta}_q h^{*\theta}_q(\hat{b_{kj}}) - h^{*\theta}_p h^{*\phi}_q(\hat{b_{kj}})
    \end{bmatrix}
\end{align}
}%

We must now relate these Jones matrix terms to the CST electric field and directivity data products, which contain, for all directions, the far-field electric field response in units [$\frac{V}{m}$], at a reference distance $1m$ away from the Vivaldi feed terminals. To be clear, the basis vectors output by CST for electric field and directivity data products are only ($\hat{\phi}$) and ($\hat{\theta}$), because ($\hat{r}$) is fixed at $1m$ for all points. For both feed polarizations, ($p,q$), the CST-calculated electric field beam may be considered as a 2x2 matrix, 

\begin{equation}\label{eq:E_Matrix}
\textbf{E}(\hat{s})=
\begin{bmatrix}
E^\phi_p & E^\theta_p\\
E^\phi_q & E^\theta_q
\end{bmatrix}
\end{equation}

Standard antenna textbook references, such as Section 16.5 of \citet{ORFANIDIS}, relate per-polarization components of the effective height matrix, $\vec{h}_p(\hat{s})$, to the per-polarization gain of the antenna, $G_p(\hat{s})$, where both $\vec{h}$ and $G_p$ are functions of $\theta$ and $\phi$.

\begin{equation}\label{eq:h_to_gain_orfanidis}
    \frac{\eta_0 \vert\vec{h}_p(\hat{s})\vert^2}{4 R_{ant}}=\frac{\lambda^2 G_p(\hat{s})}{4\pi}
\end{equation}

In our case, we seek to relate effective height to directivity, rather than gain, because directivity is another data product which is output by CST. It is well understood that gain is related to directivity by the relationship $G_p(\hat{s}) = e_{rad}D_p(\hat{s})$ , where $e_{rad}$ is the radiation efficiency. Plugging this relationship into (\ref{eq:h_to_gain_orfanidis}) and re-ordering terms,

\begin{equation}\label{eq:h_to_gain}
    \sqrt{ \frac{\eta_0}{R_{ant}} }|\vec{h}_p(\hat{s})| = \sqrt{ \frac{\lambda^2 e_{rad}D_p(\hat{s})}{\pi} }
\end{equation}

To be useful to the coupling formalism we use to make simulated visibilities, which is presented in equation (\ref{eq:V_ij_1_with_0_simp}), we must relate the direction-dependent effective height terms of (\ref{eq:JJ_Matrix}) and (\ref{eq:Jdagger_Jdagger_Matrix}) to directivity - in the direction of feed-to-feed scattering, which is the horizon plane ($\hat{s} = \hat{b_{ki}}$). 

\begin{equation}\label{eq:h_and_E_direction}
\hat{b_{ki}} = \frac{\vec{h}_p(\hat{b_{ki}})}{|\vec{h}_p(\hat{b_{ki}})|} = \frac{\vec{E}^p(\hat{b_{ki}})}{|\vec{E}^p(\hat{b_{ki}})|}
\end{equation}

Multiplying both sides of (\ref{eq:h_to_gain}) by (\ref{eq:h_and_E_direction}), we derive the non-general result

\begin{equation}\label{eq:h_to_E}
\sqrt{ \frac{\eta_0}{R_{ant}} } \vec{h}_p(\hat{b_{ki}})=\frac{ \vec{E}^p(\hat{b_{ki}}) }{ |\vec{E}^p(\hat{b_{ki}})| } \sqrt{ \frac{\lambda^2 e_{rad}D_p(\hat{b_{ki}})}{\pi} }
\end{equation}

where $\small{ |E_p(\hat{b_{ki}})| = \sqrt{|E^\phi_p(\hat{b_{ki}})|^2 + |E^\theta_p(\hat{b_{ki}})|^2} }$.

Recall that the CST far-field electric field response, $\vec{E}^p(\hat{b_{ki}})$, is calculated at a reference distance of 1m away from the feed terminals. When we plug this reference value into our impedance-normalized Jones matrix terms, e.g. ${\small \sqrt{\frac{\eta_0}{R_{ant}}}h^\phi_p(\hat{b_{ki}}) }$ or ${\small \sqrt{\frac{\eta_0}{R_{ant}}}h^\theta_q(\hat{b_{ki}}) }$, as calculated by equation (\ref{eq:h_to_E}), all factors of $R_{ant}$ and $\eta_0$ cancel in the coupling formalism as presented in equation (\ref{eq:V_ij_1_with_0_simp}). Thus, while the coupling formalism is written in terms of antenna impedance, the CST data products used in our simulations do not need to be corrected for antenna impedance. 

\bsp	
\label{lastpage}
\end{document}